\documentclass[3p,authoryear,12pt]{elsarticle}

\usepackage{graphics}
\usepackage{graphicx}	
\usepackage{amsmath}	
\usepackage{amssymb}	
\usepackage{amsthm}

\usepackage{multirow}	
\usepackage{makecell}	
\usepackage{pdflscape}	
\usepackage{url}
\usepackage{fixltx2e}
\usepackage{subfig}	
\usepackage[normalem]{ulem}

\usepackage{array}
\newcolumntype{C}[1]{>{\centering\arraybackslash}m{#1}}
\def\astrobj#1{#1}

\begin{document}

\begin{frontmatter}

\title{Photometric Investigation of Hot Exoplanets: \\ TrES-3b and Qatar-1b}

\author[carc,phys]{\c{C}. P\"{u}sk\"{u}ll\"{u}}
\ead{cpuskullu@comu.edu.tr}
\author[carc,phys]{F. Soydugan}
\author[carc,phys]{A.Erdem}
\author[carc,nzealand]{E.Budding}

\address[carc]{Astrophysics Research Centre and Observatory, \c{C}anakkale
 Onsekiz Mart University, Terzio\u{g}lu Kamp\"{u}s\"{u}, TR-17020,
 \c{C}anakkale, Turkey}

\address[phys]{Department of Physics, Faculty of Arts and Sciences,
 \c{C}anakkale Onsekiz Mart University, Terzio\u{g}lu
 Kamp\"{u}s\"{u}, TR-17020, \c{C}anakkale, Turkey}

\address[nzealand]{Carter Observatory and SCPS and
Victoria University of Wellington and Dept. Physics
\& Astronomy of UoC, New Zealand}

\begin{abstract}
New photometric follow-up observations of transitting `hot Jupiters' \astrobj{TrES-3b} and
\astrobj{Qatar-1b} are presented. Weighted mean values of the solutions of light curves in \textit{R}-filter
for both planetary systems are reported and compared with the previous results.
The transit light curves were analysed using the \textsc{winfitter} code. The physical properties of the planets were estimated.
The planet radii are found to be $R_p = 1.381 \pm 0.033  R_J$ for \mbox{\astrobj{TrES-3b}} and
$R_p = 1.142 \pm 0.025  R_J$ for \mbox{\astrobj{Qatar-1b}}.
Transit times and their uncertainties were also determined and a new linear ephemeris was computed for both systems.
Analysis of transit times showed that a significant signal could not be determined for \mbox{\astrobj{TrES-3b}},
while weak evidence was found for \mbox{\astrobj{Qatar-1b}}, which might be tested using more precise future transit times.
\end{abstract}


\end{frontmatter}

\section{Introduction}
\label{section1}

The transit technique is one of the most powerful and effective methods for discovering exoplanets.
 The number of transiting exoplanets discovered has been increasing rapidly and has now exceeded 
one thousand\footnote{Based on Transiting Extrasolar Planet Catalogue (TEPCat): \\ \url{http://www.astro.keele.ac.uk/jkt/tepcat}}.
Many of them are close-in gas giant planets also known as hot Jupiters.
The basic astrophysical parameters of the exoplanets detected by this technique, and those of their host stars,
can be derived with higher precision than those of exoplanets discovered by other methods 
\citep[e.g.][]{Southworth2008MNRAS.386.1644, Torres2008APJ.677.1324, Southworth2012.282.131, Ciceri2015AAP.577.54}.
The data quality is important in this respect. 
The photometric data collected by the \textit{Kepler} and \textit{CoRoT} satellites has allowed us 
to discover many extrasolar planets and to derive the parameters of planet-star systems, precisely. 
In addition to space missions, ground-based photometric surveys keeping eyes on whole sky, 
play very important role in discovering new hot Jupiters.
Follow up observations by the ground-based telescopes are publishing in the online database Exoplanet 
Transit Database (ETD) by many observers \citep{Poddany2010NA.15.297}. 
This cumulated data make searching for additional planets in a system possible by using transit-timing variations (TTV). 
In this study, we focus on follow-up photometric transit observations and their analysis for
hot Jupiters \astrobj{TrES-3b} and \astrobj{Qatar-1b}.

The \mbox{TrES-3b} system, discovered by \citet{O'Donovan2007APJ.663.37}, consists of a slightly metal-poor G-type star with a mass of $0.9 M_{\odot}$
and a hot Jupiter with an orbital period of 1.3 days. In their study, analysis of both photometric and radial velocity data was presented.
The planet-star system, exhibiting a v-shape transit light curve, has interested many researchers
\citep[e.g.][]{Gibson2009APJ.700.1078, Sozzetti2009APJ.691.1145, Colon2010MNRAS.408.1494, Southworth2011MNRAS.417.2166, Kundurthy2013APJ.764.8, Vavnko2013MNRAS.432.944}.
\citet{Sozzetti2009APJ.691.1145} re-determined the stellar parameters which are listed in Table~\ref{table1} and stated that the precision and accuracy of
basic planet properties may vary according to the methodologies used for analysis, as discussed by \citet{Torres2008APJ.677.1324, Rhodes2014APSS.351.451}.
\citet{Fressin2010APJ.711.374} revealed that the orbit of \mbox{TrES-3b} is circular and ruled out tidal heating from the ongoing
orbital circularization as an explanation for the inflated radius of \mbox{TrES-3b}. In order to search for additional planets in the system,
transit timing variations based on follow-up observations were investigated in several studies \citep[e.g.][]{Sozzetti2009APJ.691.1145, Christiansen2011APJ.726.94, Kundurthy2013APJ.764.8, Vavnko2013MNRAS.432.944}.
However, these authors could not find clear proof of the presence of a second planet in the system so far.

Qatar-1b has a mass of $1.09 M_J$. It was discovered by \citet{Alsubai2011MNRAS.417.709}.
The planet is revolving around its parent star in a circular orbit with a period of 1.4 days.
The host star is a slightly metal-rich K-type star with a mass of $0.85 M_{\odot}$.
Its main properties are given in Table~\ref{table1}. Later, photometric observations and
their analysis were reported by several authors
\citep[e.g.][]{Maciejewski2015AAP.577.109, Mislis2015MNRAS.448.2617, Collins2015ARXIV}.
\citet{Covino2013AAP.554.28} improved the orbital parameters and mentioned that
the host star is moderately magnetodynamically active.
A sinusoidal variation in the long-term light curves due to possible magnetic activity on the surface of the parent star
was calculated by \citet{Mislis2015MNRAS.448.2617}.
First TTV analysis presented by \citet{vonEssen2013AAP.555.92} and they indicated a significant signal with a period of $\sim190$ days.
However more recently, \citet{Maciejewski2015AAP.577.109} and \citet{Collins2015ARXIV} could not find
evidence of an additional planet in the Qatar-1b system based on TTV analysis.

The main physical properties of the host star of \mbox{TrES-3b} and \mbox{Qatar-1b} are listed in Table~\ref{table1}.
In this study, we present new transit light curves and their investigation for the exoplanets  \mbox{TrES-3b} and \mbox{Qatar-1b}.
Observations and reduction methods are given in Section~\ref{section2}. After showing parameter details emerging from
the modelling, results of the solutions are reported in Section~\ref{section3}.
The physical properties of the star-planet systems are then presented (Sect.~\ref{section4}), while updated ephemerides and TTV analysis are considered in Section~\ref{section5}.
The final section summarizes and discusses the results for these hot Jupiters in particular, in the context
of our general picture of hot Jupiter systems.

\begin{table}
	\centering
	\caption{Basic parameters of the host stars of \mbox{TrES-3b} and \mbox{Qatar-1b}.}
	\label{table1}
	\renewcommand{\arraystretch}{1.2}
	\scalebox{0.7}{
	\begin{tabular}{lcc}
		\hline
		Parameters & TrES-3 & Qatar-1 \\
		\hline
		Mass, $ M_{\star} (M_{\odot}) $ & $ 0.928_{-0.048}^{+0.028} $ & $ 0.838_{-0.041}^{+0.043} $ \\
		Radius, $ R_{\star} (R_{\odot}) $ & $ 0.829_{-0.022}^{+0.015} $ & $ 0.803 \pm 0.016 $ \\
		Density, $ {\rho}_{\star} (cgs) $ & $ 2.304 \pm 0.066 $ & $ 2.286_{-0.070}^{+0.074} $ \\
		Surface gravity, $ log\,g_{\star} (cgs) $ & $ 4.4 \pm 0.1 $ & $ 4.552_{-0.011}^{+0.012} $ \\
		Effective temperature, $ T_{eff} (K) $ & $ 5650 \pm 75 $ & $ 5013_{-88}^{+93} $ \\
		{[Fe/H]}(dex) & $ -0.19 \pm 0.08 $ & $ 0.171_{-0.094}^{+0.097} $ \\
		Age (Gyr) & $ 0.9_{-0.8}^{+2.8} $ & $ 1.19 \pm 0.47 $ \textsuperscript{*} \\
		\multirow{2}{*}{References} 	& \citet{Sozzetti2009APJ.691.1145} & \citet{Collins2015ARXIV}\\
							& & \textsuperscript{*} \citet{Mislis2015MNRAS.448.2617} \\
		\hline
	\end{tabular}
	}
\end{table}

\section{Observations and Data Reduction}

\label{section2}

Our observations were made by three different-sized telescopes and camera configurations. The 122-cm Nasymth telescope (T122) and
60-cm German-Equatorial telescope (T60) are installed at \c{C}anakkale Onsekiz Mart University Observatory (\c{C}OMUO) in Turkey.
\c{C}OMUO has a $2048 \times 2048$ pixel Apogee Alta U42 CCD (AP42) camera and a $1024 \times 1024$ pixel SBIG STL 1001E CCD (STL1001E) camera
which had been attached to the both telescopes various times.
The other telescope (T100) used in the observations is installed at T\"{U}B\.{I}TAK National Observatory (TUG) in Antalya, Turkey.
It has a primary mirror with 100-cm aperture size and is equipped with a SI 1100 CCD camera. Fields of view and pixel sizes of all telescopes and
CCD cameras are listed in Table~\ref{table2}. Photometric data of \mbox{TrES-3b} transits were taken on six nights at T122, five nights at T100 and one night at T60,
while those of \mbox{Qatar-1b} were obtained four nights each telescope. During the observations, Network Time Protocol (NTP) was used for time synchronization every minute.
T100 has miliseconds accuracy in exposure time; however T122 and T60 have accuracies in seconds.
Exposure times, CCD binning options and filters are given in Table~\ref{table3}. Depending on weather conditions, many observations
were also carried out using the defocusing method. Raw images were corrected with at least 10 biases, darks and flat images.
Photometry was performed by the \textsc{daophot} package of \textsc{iraf}\footnote{\url{http://iraf.noao.edu/}}.
We tried out different size apertures and ensured many comparison stars inside the CCD frame.
The reduction process was completed using less scattered comparisons for both \mbox{TrES-3b} and \mbox{Qatar-1b}.
Julian Date (JD\textsubscript{UTC}) was converted to Barycentric Julian Date (BJD\textsubscript{TDB}) \citep{Eastman2010PASP.122.935}.

\begin{table}
	\centering
	\caption{Specifications of telescopes and CCD cameras used in study.}
	\label{table2}
	\renewcommand{\arraystretch}{1.2}
	\scalebox{0.6}{
	\begin{tabular}{lcc}
		\hline
		Telescope and CCDs & Field of view (arcmin) & Pixel size (arcsec px\textsuperscript{-1}) \\
		\hline
		T122 + AP42 & $ 7.8 \times 7.8 $ & 0.23 \\
		T122 + STL1001E & $ 7.1 \times 7.1 $ & 0.42 \\
		T60 + AP42 & $ 19.8 \times 19.8 $ & 0.58 \\
		T60 + STL1001E & $ 18 \times 18 $ & 1.06 \\
		T100 + SI1100 & $ 21.1 \times 20.8 $ & 0.31 \\
		\hline
	\end{tabular}
	}
\end{table}

\begin{landscape}
\begin{table}
	\centering
	\caption{Journal of observations of hot Jupiters \mbox{TrES-3b} and \mbox{Qatar-1b}.}
	\label{table3}
	\renewcommand{\arraystretch}{1.2}
	\scalebox{0.8}{
	\begin{tabular}{ccccccccc}
		\hline
		Date & Telescope+CCD & Total Frame & Bin. & \makecell{Exp.Time \\ (s)} & \makecell{Airmass\\ (Beginning - Ending)} & \makecell{$\sigma_{0} $ \\ (mmag)} & $ \beta $ & \makecell{$\sigma_{s}$ \\ (mmag)} \\
		\hline
		\multicolumn{9}{c}{\mbox{TrES-3b}} \\
		03.06.2012 & 	T122+AP42 & 	178 & 	1 & 	20 & 	$ 1.02 - 1.34 $ & 	4.7 & 3.2 &	15.2 \\
		07.06.2012 & 	T122+AP42 & 	109 & 	1 & 	20 & 	$ 1.02 - 1.09 $ & 	2.8 & 0.2 &	2.8 \\
		27.05.2013 & 	T122+AP42 & 	110 & 	1 & 	100 & 	$ 1.44 - 1.01 $ & 	2.3 & 0.4 &	2.3 \\
		13.07.2013 & 	T122+AP42 & 	123 & 	1 & 	60 & 	$ 1.02 - 2.55 $ & 	2.6 & 0.4 &	2.6 \\
		15.06.2014 & 	T100+SI1100 & 	64 & 	2 & 	40 & 	$ 1.22 - 1.00 $ & 	3.4 & 0.3 &	3.4 \\
		02.07.2014 & 	T122+AP42 & 	216 & 	2 & 	45 & 	$ 1.00 - 1.08 $ & 	2.3 & 2.1 &	4.9 \\
		27.04.2015 & 	T100+SI1100 & 	196 & 	2 & 	80 & 	$ 1.98 - 1.00 $ & 	2.4 & 0.9 &	2.6 \\
		30.06.2015 & 	T122+STL1001E & 	97 & 	1 & 	80 & 	$ 1.02 - 1.32 $ & 	3.3 & 0.1 &	3.3 \\
		08.07.2015 & 	T100+SI1100 & 	58 & 	2 & 	20 & 	$ 1.06 - 1.00 $ & 	2.5 & 0.1 &	2.5 \\
		21.07.2015 & 	T100+SI1100 & 	211 & 	2 & 	80 & 	$ 1.01 - 1.92 $ & 	1.9 & 1.4 &	2.8 \\
		07.08.2015 & 	T100+SI1100 & 	154 & 	2 & 	80 & 	$ 1.00 - 1.43 $ & 	1.6 & 0.3 &	1.6 \\
		24.08.2015 & 	T60+AP42 & 	153 & 	1 & 	80 & 	$ 1.00 - 1.60 $ & 	2 & 0.9 &	3 \\
		\noalign{\smallskip}
		\hline
		\multicolumn{9}{c}{Qatar-1b} \\
		14.06.2014 & 	T100+SI1100 & 	217 & 	2 & 	60 & 	$ 1.57 - 1.14 $ & 	2.5 & 0.6 &	2.5 \\
		01.07.2014 & 	T122+AP42 & 	195 & 	1 & 	80 & 	$ 1.40 - 1.20 $ & 	2.1 & 0.8 &	2.1 \\
		28.07.2014 & 	T122+AP42 & 	261 & 	1 & 	52 & 	$ 1.13 - 1.10 - 1.21 $ & 	2.2 & 0.6 &	2.2 \\
		24.08.2014 & 	T122+AP42 & 	269 & 	1 & 	100 & 	$ 1.12 - 1.10 - 1.64 $ & 	2.5 & 0.7 &	2.5 \\
		20.09.2014 & 	T100+SI1100 & 	245 & 	2 & 	90 & 	$ 1.14 - 1.14 - 2.37 $ & 	2.2 & 0.6 &	2.7 \\
		30.09.2014 & 	T60+STL1001E & 	198 & 	1 & 	90 & 	$ 1.11 - 1.36 $ & 	2.6 & 0.9 &	2.6 \\
		11.04.2015 & 	T122+STL1001E & 	120 & 	1 & 	120 & 	$ 2.19 - 1.25 $ & 	2.2 & 0.2 &	2.2 \\
		15.05.2015 & 	T100+SI1100 & 	75 & 	2 & 	80 & 	$ 2.13 - 1.15 $ & 	3.4 & 0.1 &	3.4 \\
		25.05.2015 & 	T100+SI1100 & 	226 & 	2 & 	80 & 	$ 1.30 - 1.16 $ & 	2.3 & 0.6 &	2.6 \\
		04.08.2015 & 	T122+STL1001E & 	83 & 	1 & 	80 & 	$ 1.10 - 1.16 $ & 	2.1 & 1.2 &	2.1 \\
		03.11.2015 & 	T60+AP42 & 	135 & 	1 & 	80 & 	$ 1.25 - 1.88 $ & 	2.5 & 0.4 &	2.5 \\
		13.11.2015 & 	T60+AP42 & 	149 & 	1 & 	80 & 	$ 1.19 - 1.78 $ & 	2.1 & 0.2 &	2.1 \\
		20.11.2015 & 	T60+AP42 & 	206 & 	1 & 	80 & 	$ 1.34 - 3.00 $ & 	4.4 & 0.5 &	4.4 \\
		\hline
	\end{tabular}
	}
\end{table}
\end{landscape}

\section{Analysis of Transit Light Curves}
\label{section3}

\subsection{Method}
\label{section31}
In order to determine the physical and geometrical parameters of \mbox{TrES-3b} and \mbox{Qatar-1b} and their host stars, the \textsc{winfitter} program,
which is basically adopted from the \textsc{fitter} code, was used. \mbox{\textsc{fitter}} was originally developed by \citet{Budding1980APSS.72.369} and refined by \citet{Budding1987APJ.319.827}.
The latest user-friendly design of the program with additional features was presented by
\citet{Rhodes2014APSS.351.451}\footnote{\textsc{winfitter} can be downloaded from the website:\\ \mbox{\url{http://michaelrhodesbyu.weebly.com/astronomy.html}}}.
\textsc{winfitter} performs fitting optimization by means of a modified Marquardt-Levenberg application
to analyse the light curves of eclipsing objects. The main physical parameters of the objects,
namely, the masses ($M_{\star}$ and $M_{p}$), radii ($R_{\star}$ and $R_{p}$) and
luminosities ($L_{\star}$ and $L_{p}$) plus orbital parameters: period ($P$), orbital inclination ($i$) and
eccentricity ($e$), are used in the fitting process via a \mbox{$\chi^{2}$-minimization} computational algorithm.
In the code, correlated errors are calculated from the diagonal of the inverted  Hessian matrix.
The \textsc{fitter} code is based on the Radau model approach given by \citet{Kopal1959}.
This approach gives tidal and rotational distortions (ellipticity), together with  radiative interactions (reflection), of massive
and relatively close gravitating bodies.

We followed same analysing procedure for all transit light curves of \mbox{TrES-3b} and \mbox{Qatar-1b}. During the analysis,
the phase correction ($\Delta\phi_{0}$), reference light level ($U$), orbital inclination ($i$), ratio of radii ($k = R_{p}/R_{\star}$) and
fractional radius of the host star ($r_{1} = R_{\star}/a$ where $a$ is semi-major axis of the orbit) were taken as
free parameters. We presented some of the calculated parameters, 
$r_{2} = R_{p}/a$ and total transit duration $T_{14}$ together with literature values in the Table \ref{table4}.
The fractional value of the secondary light contribution was set to zero under the assumption that
the night side luminosity of the planet is essentially zero.
The orbit was also assumed circular. We used the \citet{Claret2011AAP.529.75} tables to calculate
the linear limb darkening coefficients: $u\textsubscript{TrES-3}$ (R) = 0.567 and  $u\textsubscript{Qatar-1}$ (R) = 0.668.
All transit curves were detrended by fitting a straight line to the out-of-transit data before analysing them in \textsc{winfitter}.
Then, we calculated the standard deviation value ($\sigma_{0}$) of this fit.
In addition to $\sigma_{0}$, in order to reveal systematic effects (red noise) on the transit curves, the time-averaged
detrending method, given by \citet{Winn2008APJ.683.1076}, was used.
As reported by \citet{Petrucci2015MNRAS.446.1389}, the red noise factor is defined as $\beta = \sigma_{r}/\sigma_{N}$, where $\sigma_{r}$ is the standard deviation of binned residuals over N points.
We selected the N numbers which represent M counts that were closer values to the division of the observation total duration by ingress/egress duration.
Therefore, $\sigma_{N}$ is the expected deviation, as given by Eq.~\ref{eq:sigmaN}:
\begin{equation}\label{eq:sigmaN}
\sigma_{N} = \frac{\sigma_{0}}{\sqrt{N}}\sqrt{\frac{M}{M-1}}
\end{equation}
Finally, the median of calculated $\beta$ values was used to scale $\sigma_{0}$ values to $\sigma_{s}$, which defines the final adopted data error, including underestimated noise (see Table~\ref{table3}).

\subsection{Solutions}
\label{section32}

\begin{figure}
	\includegraphics[width=\columnwidth]{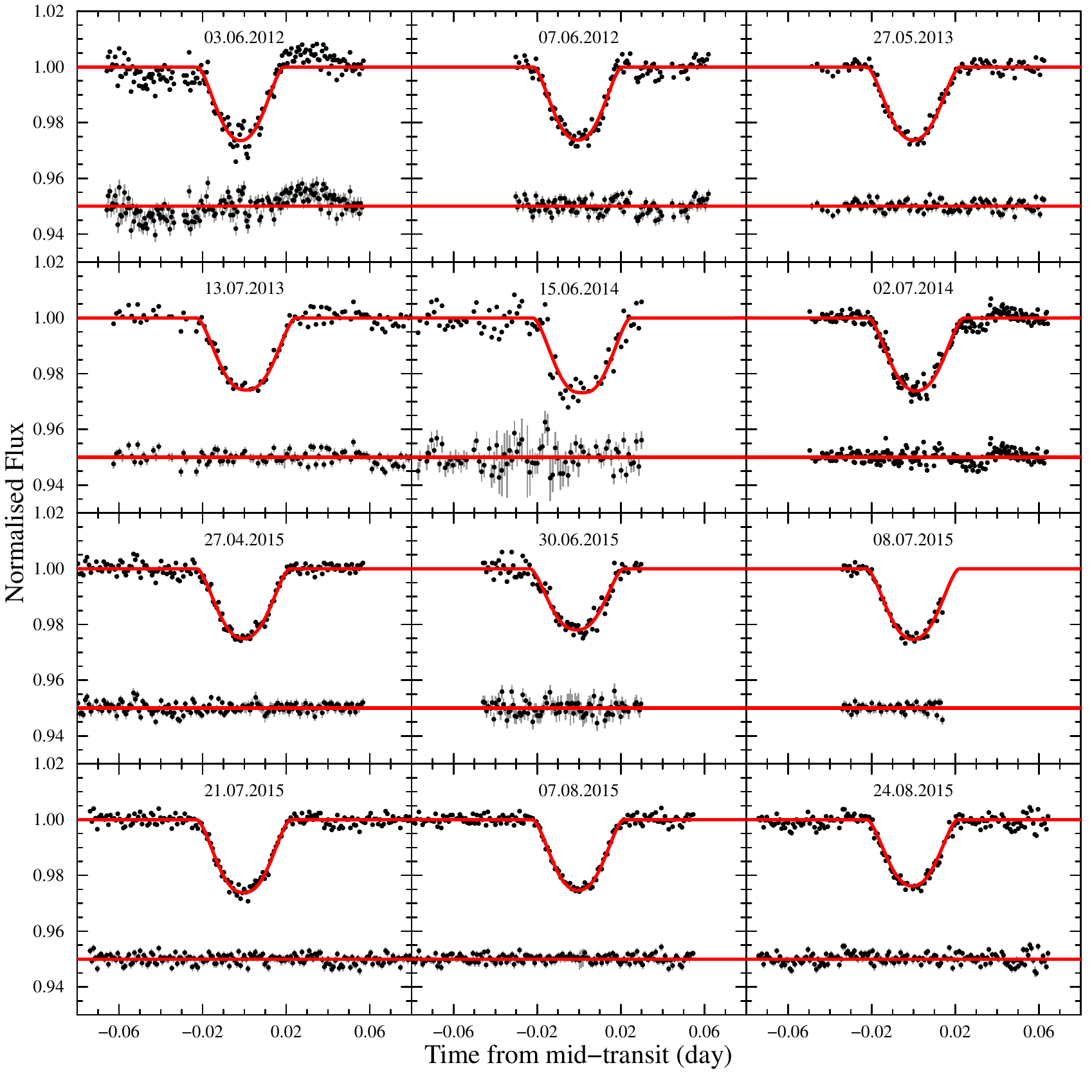}
    	\caption{Modelled individual transit curves of \mbox{TrES-3b} in \textit{R}-filter with photometric error bars and residuals from fits.}
    \label{fig1}
\end{figure}

We collected 12 transit light curves for \mbox{TrES-3b} in the \textit{R}-filter.
The red noise factors for the data taken on 3 June 2012 and 2 July 2014, are $\beta = 3.2$ and 2.1, respectively.
Therefore, these two transit light curves were eliminated from the final weighted average values.
The resulting parameters were obtained from \textsc{winfitter}, together with their errors, are listed in Table~\ref{table4},
including comparisons with previous studies.
Theoretical curves and their agreement with observational data can be seen in Fig.~\ref{fig1}. The \mbox{TrES-3b} transit model curves were found to be in the range, $\chi^{2}_{red} = 0.8 - 1.2$
over our first estimated error, $\sigma_{0}$. Hence, we cannot say definitely that there may be additional physical effects, such as maculation, i.e. spots on the surface of the host star.
The ratio of percentage change of $r_1$ is 15\%, from 0.15 to 0.18; for $k$ is 21\%, and for $i$ is 1\%, when errors are excluded.

\begin{figure}
	\includegraphics[width=\columnwidth]{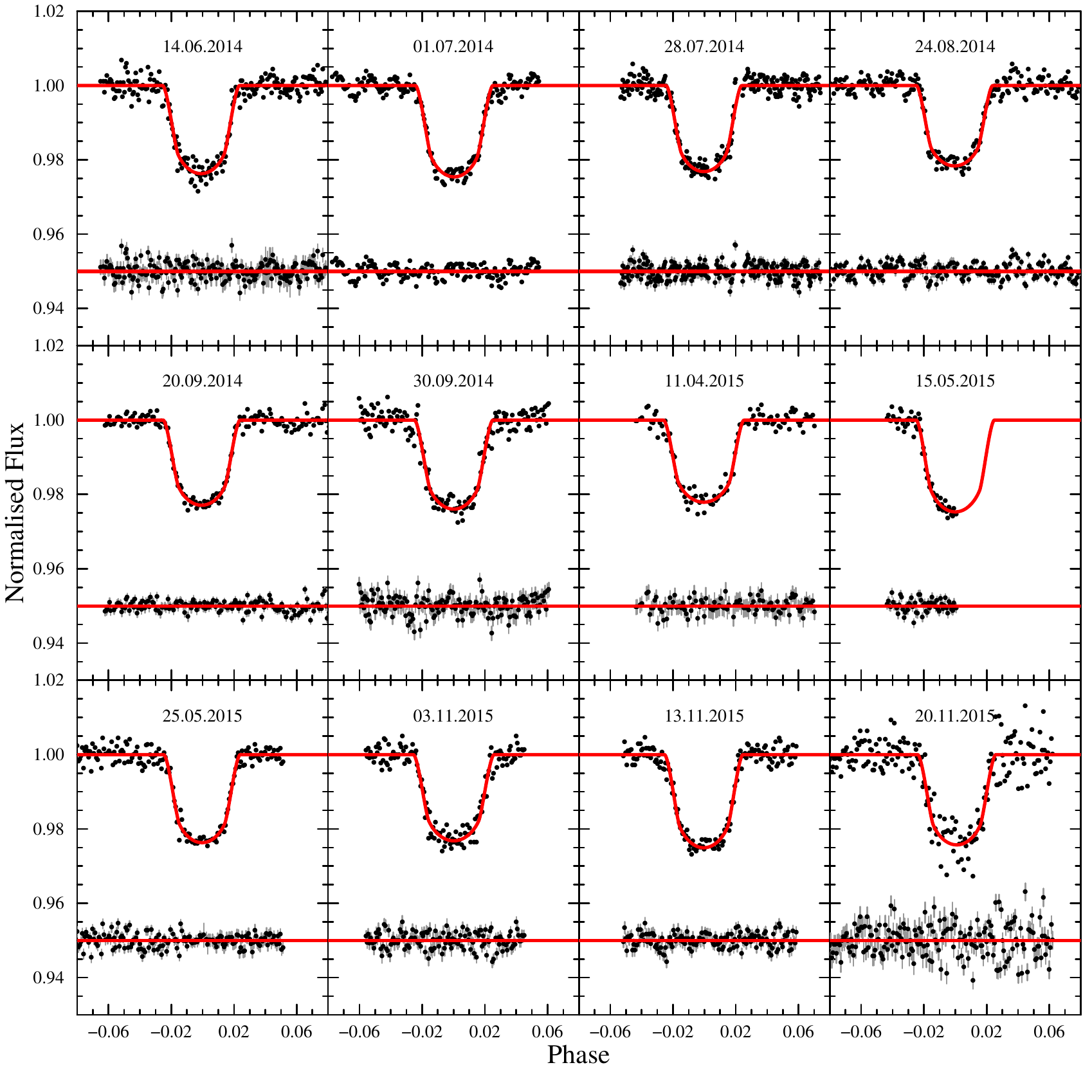}
    	\caption{Modelled individual transit curves of \mbox{Qatar-1b} in \textit{R}- filter with photometric error bars and residuals from fits.}
    \label{fig2}
\end{figure}

We obtained 12 transit light curves in the \textit{R}-filter for \mbox{Qatar-1b}.
The light curve observed on 20th November 2015 was not used in the analysis since it had large scatter, $\sigma_{0} = 4.4$
 However, the transit time was calculated for these data and evaluated for transit time analysis,
as presented in Section~\ref{section5}. The highest red noise factor for the 11 light curves was 1.2.
Weighted mean values of the resulting parameters are given in Table~\ref{table5} together with errors,
and also those presented in previous studies. Figure~\ref{fig2} shows the agreement between
the theoretical curves and transit data for \mbox{Qatar-1b}.
Since the our data scattering of \mbox{Qatar-1b} was lower than \mbox{TrES-3b},
fitting results were also derived inside a narrow parameter space which is mostly spread around 10\% for individual parameters.

Lastly, the mean system geometries were constructed using binned data and a model calculated from the weighted mean values,
as seen in Fig.~\ref{fig3}, for the star-planet systems \mbox{TrES-3b} and \mbox{Qatar-1b}. In order to check for limb darkening values,
the binned transit light curves  were solved taking limb darkening parameters as free.
These were obtained as \mbox{$u\textsubscript{TrES-3} = 0.62 \pm 0.03$} and \mbox{$u\textsubscript{Qatar-1} = 0.67 \pm 0.03$}.
These results may be compared with the theoretical values of (\mbox{$u\textsubscript{TrES-3} = 0.567$} and \mbox{$u\textsubscript{Qatar-1} = 0.668$}) by \citet{Claret2011AAP.529.75}.

\begin{table}
	\centering
	\caption{Fitting model parameters of \mbox{TrES-3b} comparing with previous studies. First column is listing references,
	following columns are the fractional radius of host star $r_{1}$, ratio of radii ($k$), orbital inclination ($i$) and total transit time ($T_{14}$) respectively.}
	\label{table4}
	\renewcommand{\arraystretch}{1.2}
	\scalebox{0.9}{
	\begin{tabular}{lcccc}
		\hline
		References & $ r_1 $ & $ k\,\,(=r_2/r_1) $ & $ i\,\,(deg) $ & $T_{14}$ (min) \\
		\hline
		\citet{O'Donovan2007APJ.663.37}	& $ 0.1650 \pm 0.0027 $ 	& $ 0.1660 \pm 0.0024 $ & $ 82.15 \pm 0.21 $ & --- \\
		\citet{Sozzetti2009APJ.691.1145} 	& $ 0.1687 \pm 0.0140 $ 	& $ 0.1655 \pm 0.0020 $ & $ 81.85 \pm 0.16 $ & --- \\
		\citet{Gibson2009APJ.700.1078} 	& --- 				& $ 0.1664 \pm 0.0011 $ & $ 81.73 \pm 0.13 $ & $ 79.92 \pm 1.44 $ \\
		\citet{Colon2010MNRAS.408.1494} 	& --- 				& $ 0.1662 \pm 0.0046 $ & --- & $ 83.77 \pm 1.15 $\\
		\citet{Southworth2010MNRAS.408.1689}\textsuperscript{a} & $ 0.1666 \pm 0.0017 $ & $ 0.1639 \pm 0.0037 $ & $ 82.07 \pm 0.17 $ & --- \\
		\citet{Lee2011PASJ.63.301}       & $ 0.1674 \pm 0.0023 $ 	& $ 0.1603 \pm 0.0042 $ & $ 81.77 \pm 0.14 $ & --- \\
		\citet{Christiansen2011APJ.726.94} & $ 0.1664 \pm 0.0204 $ 	& $ 0.1661 \pm 0.0343 $ & $ 81.99 \pm 0.30 $ & $ 81.9 \pm 1.10 $ \\
		\citet{Southworth2011MNRAS.417.2166}\textsuperscript{a} & $ 0.1682 \pm 0.0014 $ & $ 0.1635 \pm 0.0025 $ & $ 81.93 \pm 0.13 $ & --- \\
		\multirow{2}{*}{\citet{Kundurthy2013APJ.764.8}\textsuperscript{b}}
			& $ 0.1675 \pm 0.0008 $ & $ 0.1652 \pm 0.0009 $ & $ 81.95 \pm 0.06 $ & --- \\
			& $ 0.1698 \pm 0.0014 $ & $ 0.1649 \pm 0.0015 $ & $ 81.51 \pm 0.14 $ & --- \\
		\citet{Turner2013MNRAS.428.678} & $ 0.1721 \pm 0.0054 $ 	& $ 0.1693 \pm 0.0087 $ & $ 81.35 \pm 0.63 $ & $ 81.3 \pm 0.23 $ \\
		\multirow{2}{*}{\citet{Vavnko2013MNRAS.432.944}\textsuperscript{b}}
			& $ 0.1682 \pm 0.0032 $ & $ 0.1644 \pm 0.0047 $ & $ 81.86 \pm 0.28 $ & $ 79.2 \pm 1.38 $ \\
			& $ 0.1696 \pm 0.0024 $ & $ 0.1669 \pm 0.0027 $ 	& $ 81.76 \pm 0.14 $ 	& $ 79.08 \pm 0.72 $ \\
		This study & $ 0.1691 \pm 0.0024 $ 	& $ 0.1709 \pm 0.0030 $ 	& $ 81.83 \pm 0.21 $ & $ 83.98 \pm 0.75 $ \\
		\hline
		\multicolumn{5}{l} {\textsuperscript{a} present solutions obtained from light curves in the literature.} \\
		\multicolumn{5}{l} {\textsuperscript{b} present two solutions estimated using different methods.}
	\end{tabular}
	}
\end{table}

\begin{table}
	\centering
	\caption{Fitting model parameters of \mbox{Qatar-1b} comparing with previous studies.
	The column parameters are same as in the Table \ref{table4}.}
	\label{table5}
	\renewcommand{\arraystretch}{1.2}
	\scalebox{0.9}{
	\begin{tabular}{lcccc}
		\hline
		References & $ r_1 $ & $ k\,\,(=r_2/r_1) $ & $ i\,\,(deg) $ & $T_{14}$ (min) \\
		\hline
		\citet{Alsubai2011MNRAS.417.709}	& $ 0.1633 \pm 0.0053 $ & $ 0.1454 \pm 0.0015 $ & $ 83.47 \pm 0.40 $ & $ 96.71 \pm 1.11 $ \\
		\citet{vonEssen2013AAP.555.92}	& $ 0.1558 \pm 0.0024 $ & $ 0.1435 \pm 0.0008 $ & $ 84.52 \pm 0.24 $ & --- \\
		\citet{Covino2013AAP.554.28}	& $ 0.1601 \pm 0.0025 $ & $ 0.1513 \pm 0.0008 $ &	$ 83.82 \pm 0.25 $ & $ 97.63 \pm 1.44 $ \\
		\citet{Mislis2015MNRAS.448.2617} 	& $ 0.1640 \pm 0.0030 $ & $ 0.1475 \pm 0.0009 $ & $ 84.03 \pm 0.16 $& --- \\
		\citet{Maciejewski2015AAP.577.109} & $ 0.1582 \pm 0.0017 $ & $ 0.1459 \pm 0.0008 $ &	$ 84.26 \pm 0.17 $ & $ 98.50 \pm 1.70 $ \\
		\citet{Collins2015ARXIV} 	& $ 0.1600 \pm 0.0018 $ & $ 0.1463 \pm 0.0006 $ & $ 84.08 \pm 0.16 $ & $ 99.66 \pm 0.47 $ \\
		This study		& $ 0.1584 \pm 0.0031 $ & $ 0.1470 \pm 0.0012 $ &	$ 84.42 \pm 0.30 $ & $ 100.84 \pm 0.92 $ \\
		\hline
	\end{tabular}
	}
\end{table}

\section{Physical Properties}
\label{section4}

\begin{figure}
	\centering
	\includegraphics[width=\columnwidth]{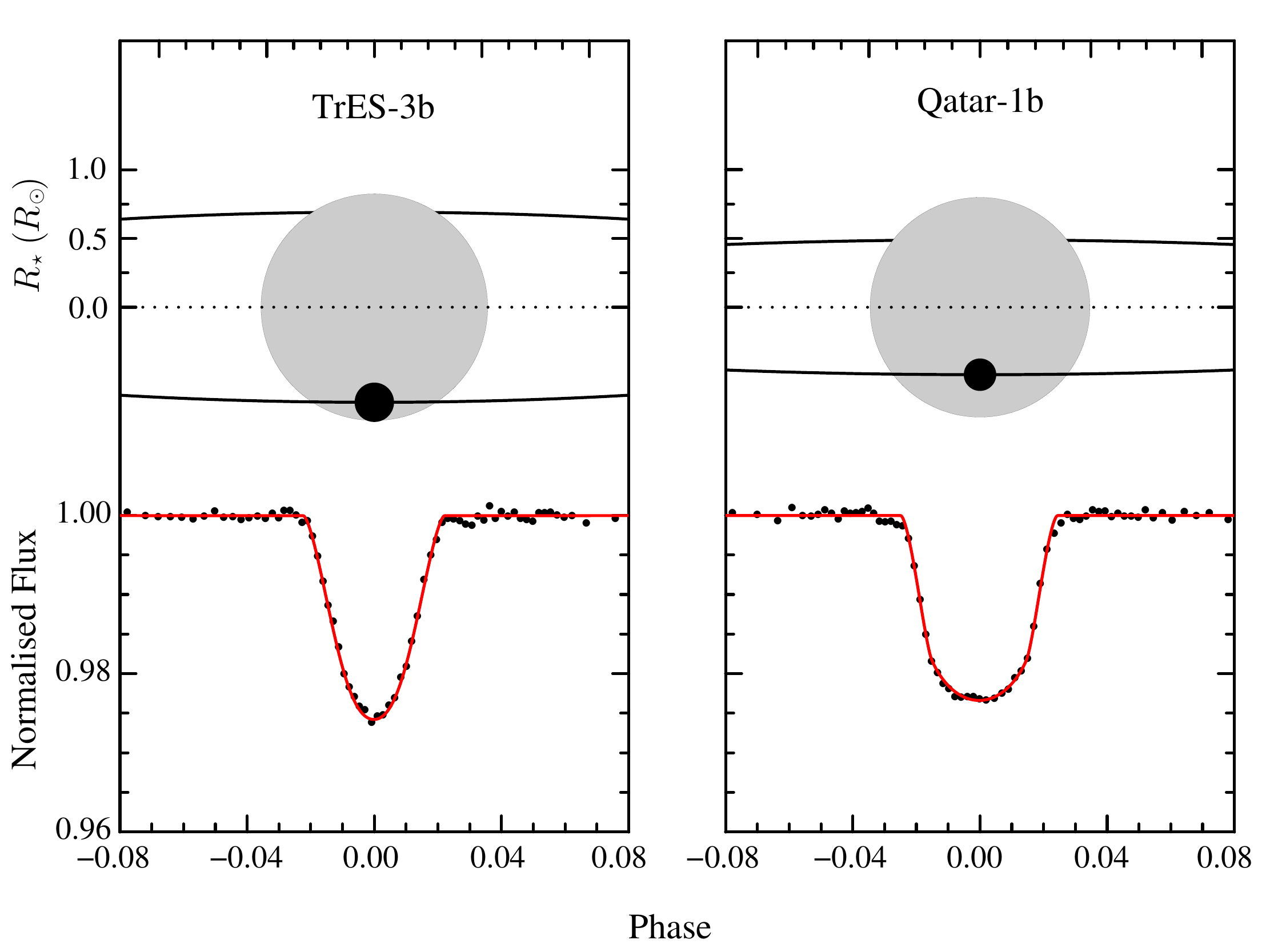}
	\caption{System geometries and transit light curves at zero point phase both TrES-3b (left panel) and Qatar1b (right panel).
   	The observational points indicate combined transit light curve of the planet for all normalized individual transit curves binned to 70 points.
    	The continuous line shows the theoretical curve were calculated using the weighted mean solution.}
	\label{fig3}
\end{figure}

In order to determine the astrophysical parameters of the hot Jupiters, \mbox{TrES-3b} and \mbox{Qatar-1b},
the  photometric results in Table~\ref{table5} together with basic parameters given in Table~\ref{table1} were used.
The mass of planet and its host star in the \mbox{Qatar-1} system were taken from \citet{Alsubai2011MNRAS.417.709}
while the mass values of \mbox{TrES-3b} and its host star were used as given by \citet{Sozzetti2009APJ.691.1145}
and \citet{Torres2008APJ.677.1324}, respectively.
We calculated the physical parameters of the stars ($ R_{\star}, \rho_{\star} $, log$\,g_{\star}, T_{eff}$) and
planets ($R_p, \rho_p,$ log$\,g_p, T{'}_{eq}$) from the weighted mean values of all transit parameters.
These results with standard errors are given in Table~\ref{table6}.
A comparison between the binned and combined transit data, and theoretical light curves were calculated from
the weighted mean parameters listed in Tables~\ref{table4} and~\ref{table5}, can be seen in Fig.~\ref{fig3}
for \mbox{TrES-3b} and \mbox{Qatar-1b}, respectively. Binning was done to 70 points for each transit curve.
Transit geometries are also plotted with the corresponding curves (see upper parts of the Figures~\ref{fig3}).

As a measure of gravitational focusing, the Safronov number ($\Theta$) given by \citet{Hansen2007APJ.671.861} is
\begin{equation}\label{eq:safronov}
\Theta=\frac{a}{R_p}\frac{M_p}{M_{\star}}.
\end{equation}

They proposed a classification for exoplanets using the values of the Safranov number. According to this designation,
exoplanets with \mbox{$\Theta \sim 0.07 \pm 0.01$} were termed Class I, while those having $\Theta \sim 0.04 \pm  0.01$ were Class II.
While we calculated \mbox{$\Theta\textsubscript{TrES-3 b} = 0.068 \pm 0.003$} for \mbox{TrES-3b}, meaning that it falls within the Class I group, \mbox{Qatar-1b} appears to be
between Class I and Class II, with \mbox{$\Theta\textsubscript{\mbox{Qatar-1b}} = 0.053 \pm 0.002$}.

\begin{figure}
	\centering
	\includegraphics[width=10cm]{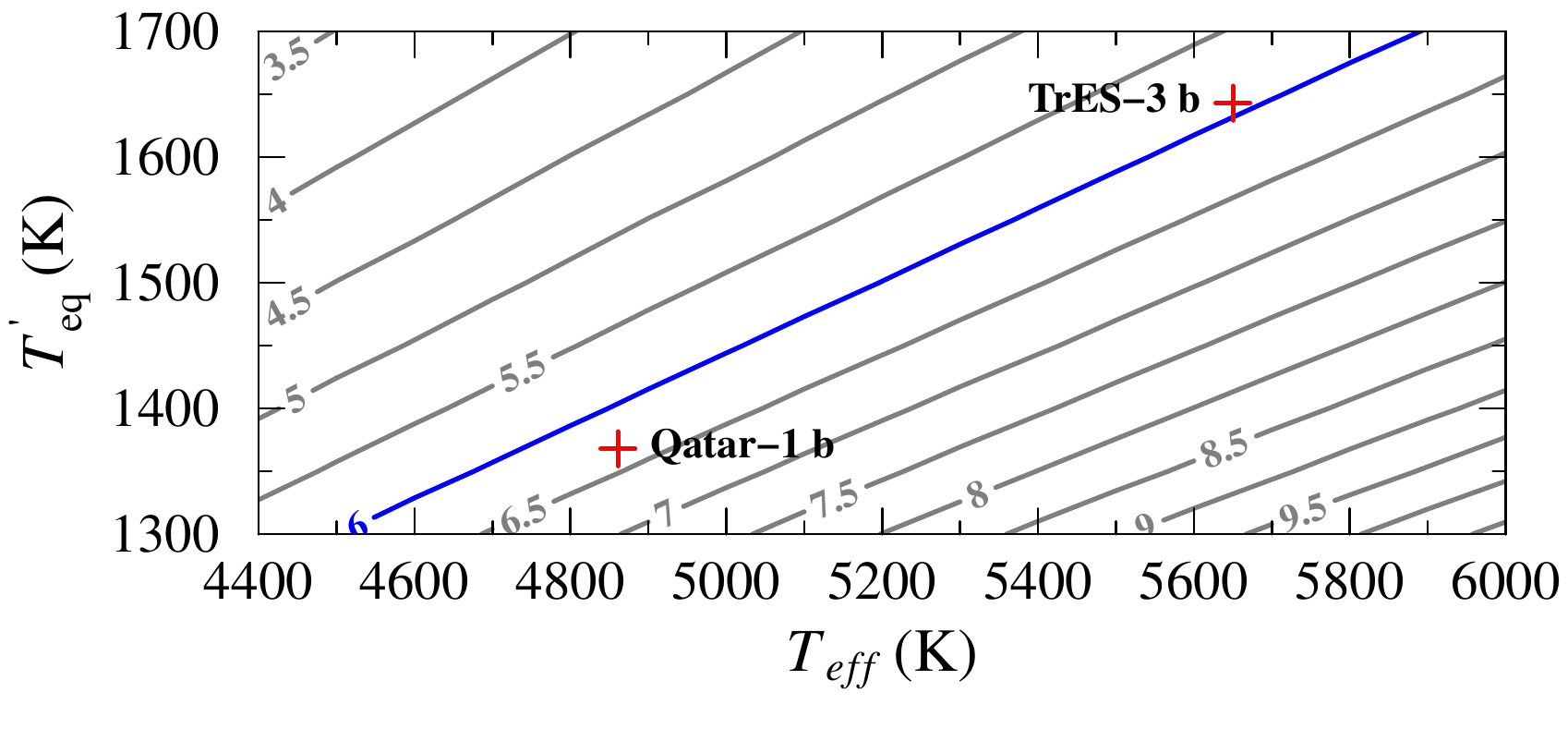}
    	\caption{Temperature parameters, $T_{eff} $ and $ T_{eq}^{'}$ plotted together in order to view diversity between
    	flux received on planets by their host stars. Solid lines indicate $a/R_{\star}$ steps. The planets are labelled with crossmarks.
    	Both planets, as expected, appear on same line, of $a/R_{\star} \sim 6$, coloured blue.}
    \label{fig5}
\end{figure}

The equilibrium temperature was also estimated by the simplified equation given by \citet{Southworth2010MNRAS.408.1689}:
\begin{equation}\label{eq:teq}
\ T_{eq}^{'}=T_{eff}\sqrt{\frac{R_{\star}}{2a}}
\end{equation}
where $a$ is in same unit with stellar radius, $R_{\star}$. The effective temperature of the host stars,
$T_{eff}$ were obtained from \citet{Torres2008APJ.677.1324} and \citet{Alsubai2011MNRAS.417.709}
for \mbox{TrES-3b} and \mbox{Qatar-1b}, respectively.
Due to the similar-size orbit and almost identical $R_{\star}$ values, the ratio $T_{eq}^{'}/T_{eff}$ should be approximately
the same for these two exoplanets.
In order to illustrate this similarity, we plotted $T_{eff}$ against $T_{eq}^{'}$ with the semi-major axis in stellar radii,
$a/R_{\star}$ contour lines in Fig.~\ref{fig5}. As expected, both planets lie on the same $a/R_{\star}$ contour line.

\begin{table}
	\centering
	\caption{Astrophysical parameters of star-planet systems \mbox{TrES-3b} and \mbox{Qatar-1b}.}
	\label{table6}
	\renewcommand{\arraystretch}{1.2}
	\scalebox{0.8}{
	\begin{tabular}{lcc}
		\hline
		Parameters & \mbox{TrES-3b} & \mbox{Qatar-1b} \\
		\hline
		\noalign{\smallskip}
		\multicolumn{3}{l} {Adopted values} \\
		\quad Stellar mass, $M_{\star} (M_{\odot})$ 		& $ 0.928 \pm 0.031 $ & $ 0.850 \pm 0.030 $ \\
		\quad Planet mass, $M_p (M_J)$ 			& $ 1.910 \pm 0.065 $ & $ 1.090 \pm 0.080 $ \\
		\quad Effective temperature of the host star, $T_{eff} (K)$ & $ 5650 \pm 75 $ & $ 4861 \pm 125 $ \\
		\multicolumn{3}{l} {Derived values} \\
		\quad Semi-major axis, $a/R_{\star}$ 		& $ 5.92 \pm 0.11 $ & $ 6.31 \pm 0.15 $ \\
		\quad Stellar radius, $R_{\star} (R_{\odot})$ 		& $ 0.826 \pm 0.012 $ & $ 0.798 \pm 0.016 $ \\
		\quad Stellar density, $\rho_{\star} (\rho_{\odot})$ 	& $ 1.607 \pm 0.074 $ & $ 1.666 \pm 0.103 $ \\
		\quad Stellar surface gravity, $logg_{\star} (cgs)$	& $ 4.57 \pm 0.01 $ & $ 4.56 \pm 0.02 $ \\
		\quad Planet radius, $R_p (R_J)$ 			& $ 1.381 \pm 0.033 $ & $ 1.142 \pm 0.025 $ \\
		\quad Planet density, $\rho_p (\rho_J)$ 		& $ 0.668 \pm 0.049 $ & $ 0.679 \pm 0.045 $ \\
		\quad Planet surface gravity, $logg_p (cgs)$ 		& $ 3.39 \pm 0.02 $ & $ 3.32 \pm 0.02 $ \\
		\quad Equilibrium temperature of the planet, $T_{eq}^{'} (K)$ & $ 1643 \pm 25 $ & $ 1368 \pm 30 $ \\
		\quad Safronov Number, $ \Theta $ 			& $ 0.068 \pm 0.003 $ & $ 0.053 \pm 0.002 $ \\
		\quad Incident flux, $  \langle F \rangle  (10^9 erg s^{-1} cm^{-2})$ & $ 1.650 \pm 0.079 $ & $ 0.798 \pm 0.060 $ \\
		\hline
	\end{tabular}
	}
\end{table}

\section{Transit Ephemeris and Transit Timing Variations}
\label{section5}

The investigation of additional bodies or unseen components around binary stars is a common research area
in stellar astrophysics. In particular, the O-C method based on eclipse times is frequently used for this purpose.
The method is also preferred when searching for additional planets around star-planet systems.
In the field of multi-planet research, very precise transit times at least in several tens seconds of accuracy
are required since the expected amplitudes in the O-C diagrams are very low \citep[e.g.][]{Lithwick2012APJ.761.122}.

In this study, we also used the O-C method to analyse transit times of the target hot Jupiters. As a first step,
mid-transit times from \textsc{winfitter} solutions for both the \mbox{TrES-3b} and \mbox{Qatar-1b} systems were determined.
The code calculates a $\Delta\phi_{0}$ parameter, which was later converted to transit time in BJD.
We also supplemented our data by retrieving values having errors lower than 0.0005 BJD (43.2 s)
from literature and the ETD, selecting clear transit curves.
In this study, 12 mid-transit times for \mbox{TrES-3b} and 11 mid-transit times for \mbox{Qatar-1b} were obtained
and are listed in Table~\ref{table7}. Epoch numbers and O-C values in Table~\ref{table7} were calculated
using the new transit ephemeris in Eqs.~\ref{eq:t0_tres3b} and~\ref{eq:t0_qatar1b},
which led to improvements for each object, as presented below.
We generated Lomb-Scargle periodograms \citep{Lomb1976APSS.39.447, Scargle1982APJ.263.835}
inside frequency windows limited to the Nyquist frequency with the improved TTVs.
The false alarm probability (FAP) was also checked to ensure that the detected dominant frequencies were significant
\citep{Horne1986APJ.302.757}.

\begin{table}
	\centering
	\caption{Mid-transit times of \mbox{TrES-3b} and \mbox{Qatar-1b}.}
	\label{table7}
	\renewcommand{\arraystretch}{1.2}
	\scalebox{1}{
	\begin{tabular}{ccccc}
		\hline
		Date  & Filter & Epoch & $T_{0}$ (BJD 2450000+) & O-C (days) \\
		\hline
		\multicolumn{5}{c} {\mbox{TrES-3b}} \\
		03.06.2012 &	 \textit{R}\textsubscript{Bessell} &	 0 &	 $ 6082.49611 \pm 0.00056 $ &	0.00221 \\
		07.06.2012 &	 \textit{R}\textsubscript{Bessell} &	 3 &	 $ 6086.41313 \pm 0.00034 $ &	0.00067 \\
		27.05.2013 &	 \textit{R}\textsubscript{Bessell} &	 274 &	 $ 6440.38897 \pm 0.00028 $ &	-0.00004 \\
		13.07.2013 &	 \textit{R}\textsubscript{Bessell} &	 310 &	 $ 6487.41045 \pm 0.00043 $ &	-0.00127 \\
		15.06.2014 &	 \textit{R}\textsubscript{Bessell} &	 568 &	 $ 6824.40585 \pm 0.00064 $ &	-0.00198 \\
		02.07.2014 &	 \textit{R}\textsubscript{Bessell} &	 581 &	 $ 6841.38641 \pm 0.00030 $ &	-0.00185 \\
		27.04.2015 &	 \textit{R}\textsubscript{Bessel} &	 810 &	 $ 7140.50549 \pm 0.00033 $ &	0.00052 \\
		30.06.2015 &	 \textit{R}\textsubscript{Cousins} &	 859 &	 $ 7204.50938 \pm 0.00044 $ &	0.00126 \\
		08.07.2015 &	 \textit{R}\textsubscript{Bessell} &	 865 &	 $ 7212.34536 \pm 0.00037 $ &	0.00013 \\
		21.07.2015 &	 \textit{R}\textsubscript{Bessell} &	 875 &	 $ 7225.40740 \pm 0.00030 $ &	0.00030 \\
		07.08.2015 &	 \textit{R}\textsubscript{Bessell} &	 888 &	 $ 7242.38793 \pm 0.00028 $ &	0.00040 \\
		24.08.2015 &	 \textit{R}\textsubscript{Bessell} &	 901 &	 $ 7259.36845 \pm 0.00034 $ &	0.00050 \\
		\hline
		\multicolumn{5}{c} {\mbox{Qatar-1b}} \\
		14.06.2014 &	 \textit{R}\textsubscript{Bessell} &	 0 &	 $ 6823.41500 \pm 0.00044 $ &	0.00138 \\
		01.07.2014 &	 \textit{R}\textsubscript{Bessell} &	 12 &	 $ 6840.45327 \pm 0.00042 $ &	-0.00065 \\
		28.07.2014 &	 \textit{R}\textsubscript{Bessell} &	 31 &	 $ 6867.43480 \pm 0.00042 $ &	0.00041 \\
		24.08.2014 &	 \textit{R}\textsubscript{Bessell} &	 50 &	 $ 6894.41548 \pm 0.00047 $ &	0.00063 \\
		20.09.2014 &	 \textit{R}\textsubscript{Bessell} &	 69 &	 $ 6921.39618 \pm 0.00048 $ &	0.00085 \\
		30.09.2014 &	 \textit{R}\textsubscript{Cousins} & 76 &	 $ 6931.33528 \pm 0.00053 $ &	-0.00022 \\
		11.04.2015 &	 \textit{R}\textsubscript{Cousins} & 212 &	 $ 7124.45935 \pm 0.00059 $ &	0.00047 \\
		25.05.2015 &	 \textit{R}\textsubscript{Bessell} &	 243 &	 $ 7168.48014 \pm 0.00058 $ &	0.00050 \\
		03.11.2015 &	 \textit{R}\textsubscript{Bessell} &	 357 &	 $ 7330.36216 \pm 0.00065 $ &	-0.00030 \\
		13.11.2015 &	 \textit{R}\textsubscript{Bessell} &	 364 &	 $ 7340.30315 \pm 0.00064 $ &	0.00051 \\
		20.11.2015 &	 \textit{R}\textsubscript{Bessell} &	 369 &	 $ 7347.40134 \pm 0.00085 $ &	-0.00142 \\
		\hline
	\end{tabular}
	}
\end{table}

\subsection{\mbox{TrES-3b}}
\label{section51}
\begin{figure}
	\includegraphics[width=\columnwidth]{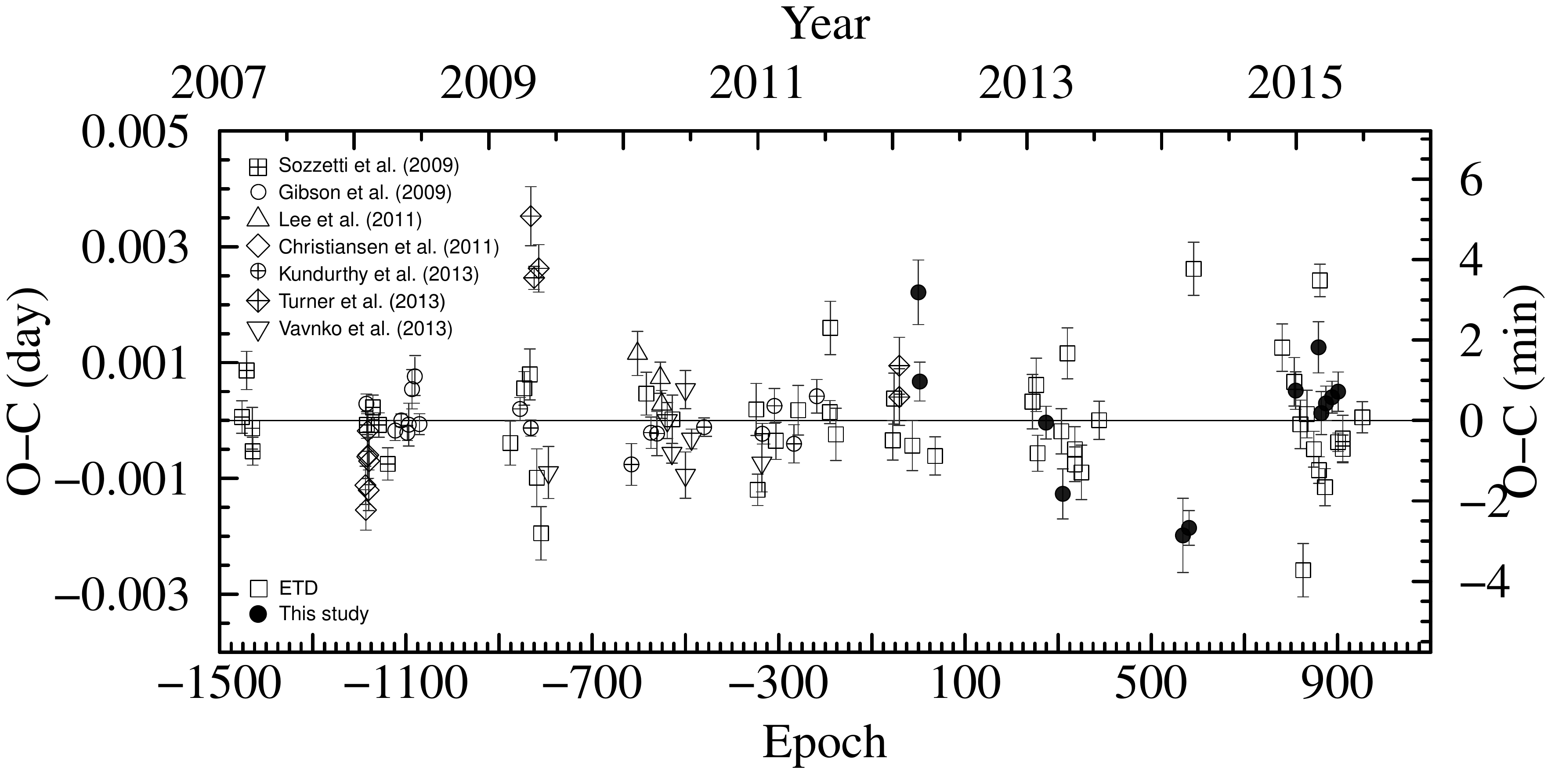}
    	\caption{O-C diagram for \mbox{TrES-3b}.}
    \label{fig6}
\end{figure}

New light elements of the \mbox{TrES-3b} were determined using our transit times together
with data from the ETD and times given by
\cite{Sozzetti2009APJ.691.1145, Gibson2009APJ.700.1078, Lee2011PASJ.63.301, Christiansen2011APJ.726.94, Kundurthy2013APJ.764.8, Turner2013MNRAS.428.678} and \citet{Vavnko2013MNRAS.432.944}.
An improved ephemeris was calculated as follows:
\begin{equation}\label{eq:t0_tres3b}
T_{c} = (BJD) \ 2456082.49390(3) + 1.30618652(4) \times E
\end{equation}

where E is the epoch number. Using the weighted linear fits to the O-C data, we obtained a reduced chi-squared value,
$\chi^{2}_{red} = 7.8$. Despite this high value of $\chi^{2}_{red}$,
no peaks in our Lomb-Scargle analyses for \mbox{TrES-3b} exceeded 5\% of the significance level.
Since no clear dominant peak was observable; we would like to present the highest peak at a frequency of
\mbox{$\nu_{1}\textsubscript{(TrES-3 b)} = 0.0154 \pm 0.0001$ \ cycl $P^{-1}$}, which corresponds to
a TTV period of $P_{ttv} = 84.79 \pm 0.35$ days. The O-C diagram of \mbox{TrES-3b} can be seen in Fig.~\ref{fig6}.

\subsection{\mbox{Qatar-1b}}
\label{section52}
\begin{figure}
	\includegraphics[width=\columnwidth]{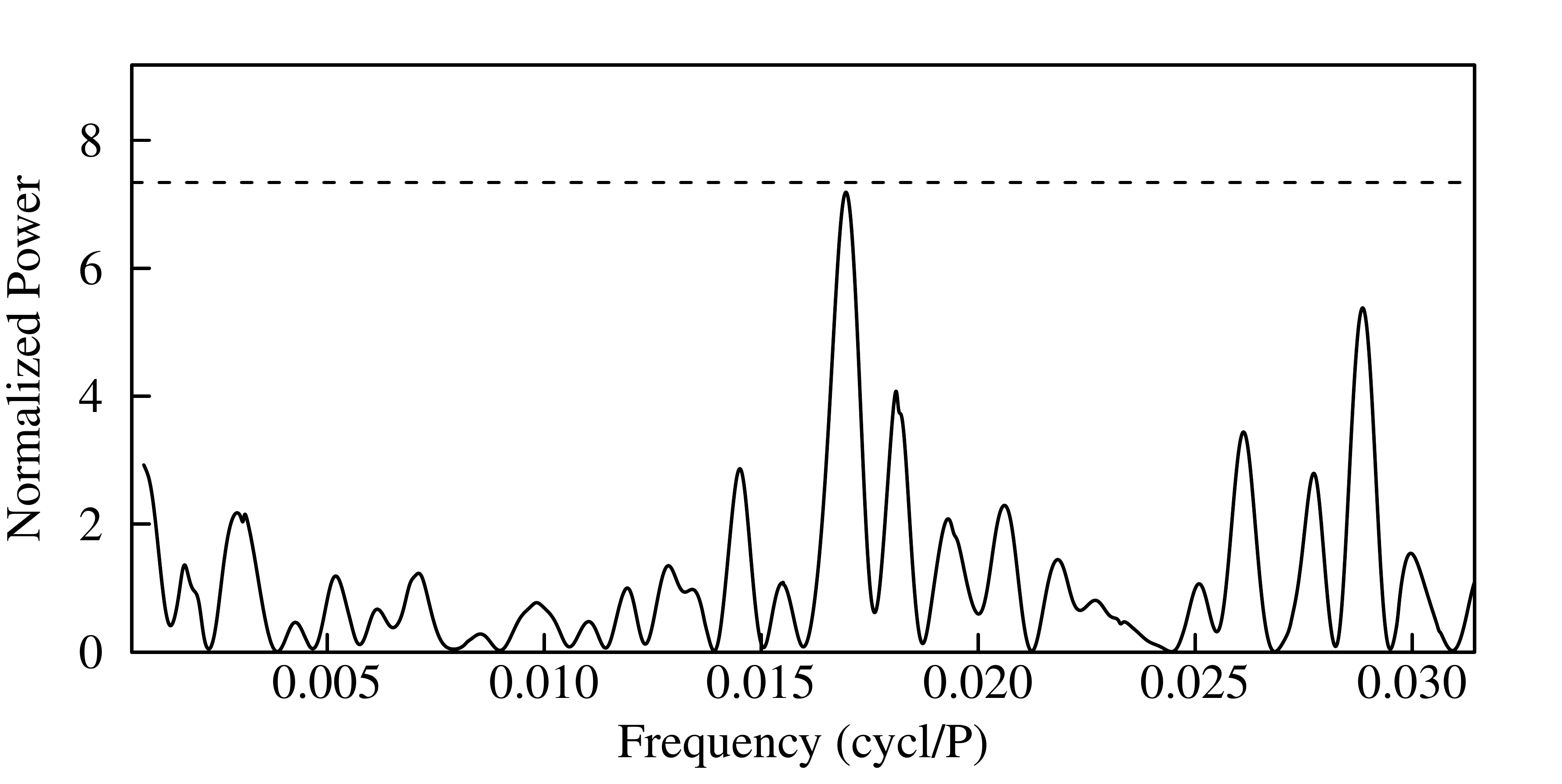}
    	\caption{Lomb-Scargle periodogram of O-C for \mbox{Qatar-1b}. Dashed line shows 5\% significance level.}
    \label{fig7}
\end{figure}

\begin{figure}
	\includegraphics[width=\columnwidth]{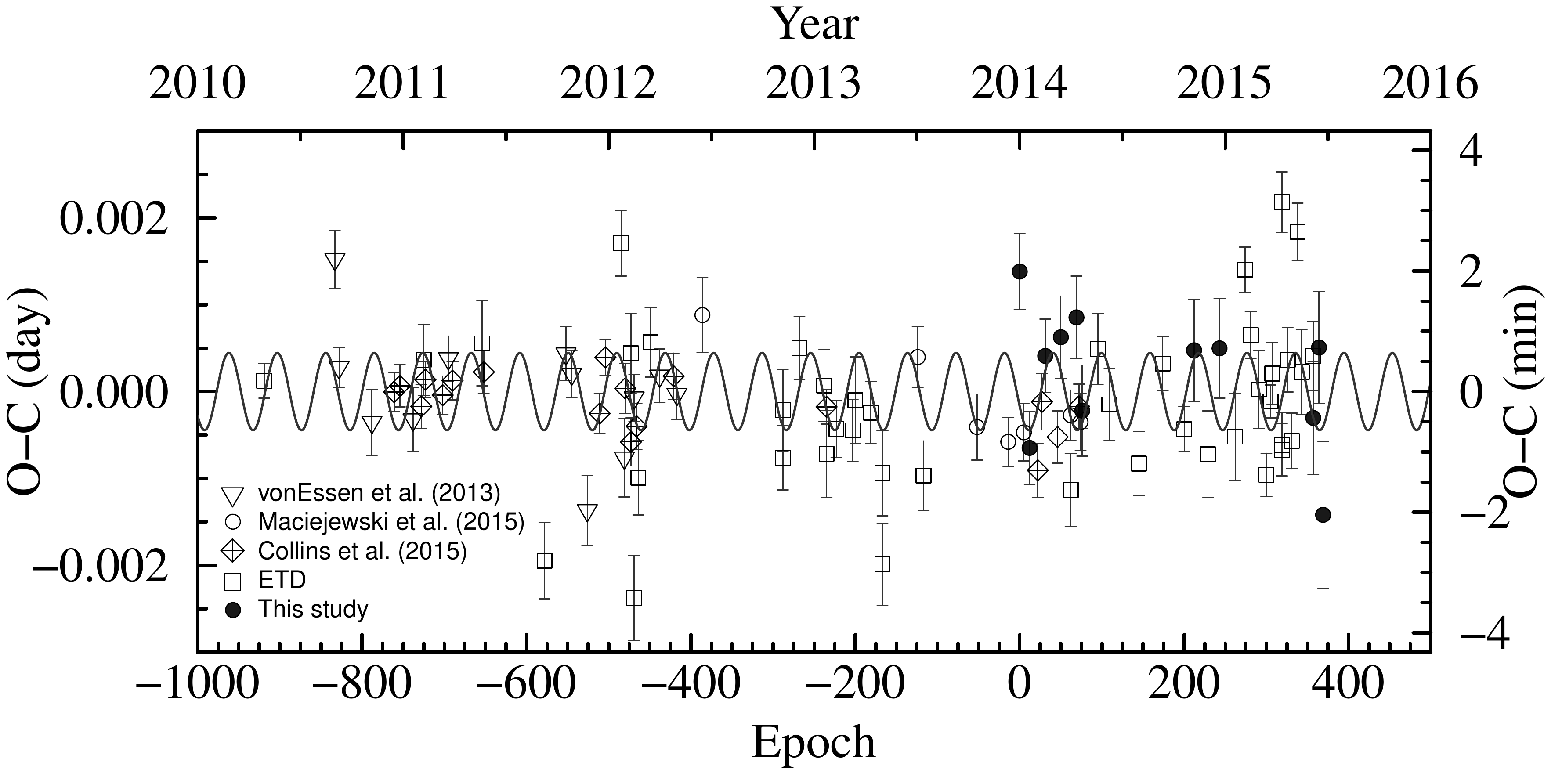}
    	\caption{O-C diagram for \mbox{Qatar-1b}.
	Applied least square fit to sinusoidal variation with obtained a reduced $\chi^{2}_{red} = 3.9$,
	is indicated by the continuous line.}
    \label{fig8}
\end{figure}

We used transit times for \mbox{Qatar-1b} was  by combining our mid-transit times with data
from \citet{vonEssen2013AAP.555.92}, \citet{Maciejewski2015AAP.577.109} and ETD.
The improved linear ephemeris obtained is as follows:
\begin{equation}\label{eq:t0_qatar1b}
T_c = (BJD) \ 2456823.41361(4) + 1.4200248(1) \times E
\end{equation}
By applying linear least squares fits, a reduced $\chi^{2}_{red} = 4.5$ was arrived at.
The corresponding Lomb-Scargle periodogram shows a few strong peaks,
but they are all under 5\% significance level.
The highest peak is seen at the frequency
\mbox{$\nu_{1}\textsubscript{(\mbox{Qatar-1b})} = 0.0169 \pm 0.0001$ \ cycl  $P^{-1}$},
which corresponds to $P_{ttv} = 83.75 \pm 0.48$ days (see Fig.~\ref{fig7}).
We obtained a value of $\chi^{2}_{red} = 3.9$  after applying a least squares sinusoidal fit to the O-C data
with a derived amplitude of $A_{ttv} = 0.00044 \pm 0.00011 $ days ($38.0 \pm 9.5$ s). 
In Fig.~\ref{fig8}, the distribution of the O-C data together with its sinusoidal TTV can be seen.
The parameters of this sinusoidal variation were found to be similar to those determined in \citet{vonEssen2013AAP.555.92}.

\section{Discussion and Conclusions}
\label{section6}
This study has given a photometric investigation of the transiting exoplanets \mbox{TrES-3b} and \mbox{Qatar-1b}.
The transit light curves covering or partly covering the transit phases for both planets, were used to determine their photometric parameters, which were found to be generally consistent with previous results.
It is of interest to examine these exoplanets together and compare their parameters, since they have
approximately similar size orbits.

\begin{figure}
	\centering
	\includegraphics[width=12cm]{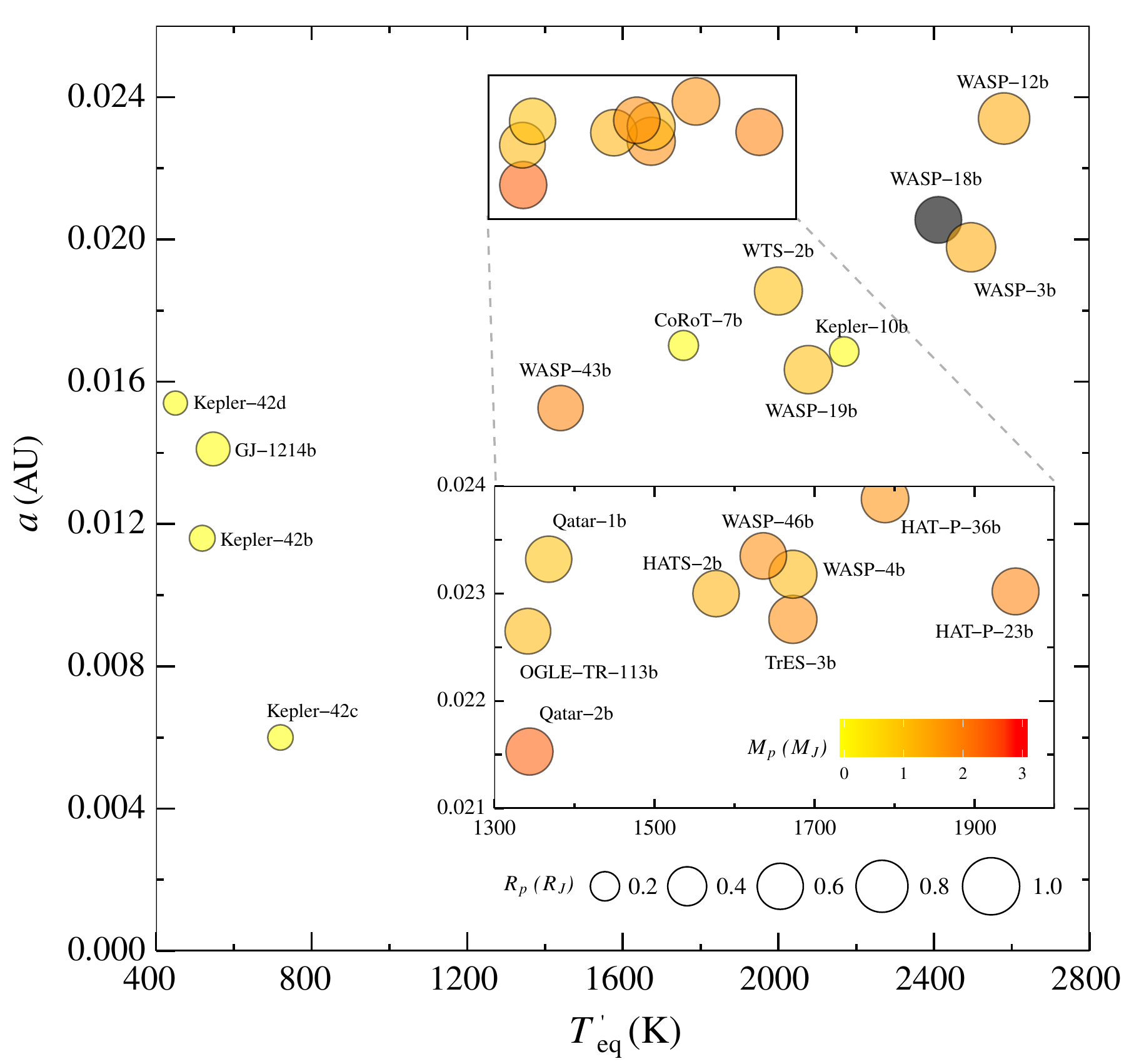}
    	\caption{Location of exoplanets having orbit size $(a)$ values smaller than $0.024$ AU in plane of $ T_{eq}^{'} - a$
	diagram. Data except for \mbox{TrES-3b} and \mbox{Qatar-1b} taken from TEPCat database.
	The super massive ($M_p = 10.5 M_J$)  planet WASP-18b coloured grey.}
    \label{fig9}
\end{figure}

We found a radius of \mbox{$R_p = 1.381 \pm 0.033 R_J$} for TrES-3b and
$R_p = 1.142 \pm 0.025 R_J$ for \mbox{Qatar-1b}.
Although there are similarities in the orbits of the planets, these planets and their host stars have
different physical properties (see Fig.~\ref{fig9}). The most conspicuous differences are in the masses of the planets
and the metallicities of their host stars. The mass of \mbox{TrES-3b} is 1.91 $M_J$ while that of \mbox{Qatar-1b} is 1.09 $M_J$.
These differences put the planets in separate classes when defined by their Safronov numbers. The more massive planet
\mbox{TrES-3b} is a Class I planet with \mbox{$\Theta\textsubscript{\mbox{TrES-3b}} = 0.068 \pm 0.002$},
while \mbox{Qatar-1b} is a Class II planet with \mbox{$\Theta\textsubscript{\mbox{Qatar-1b}} = 0.053 \pm 0.002$}.

The transit times and their uncertainties were determined and analysed for possible transit time variations (TTVs).
We obtained 12 mid-transit times for \mbox{TrES-3b} and 11 mid-transit times for \mbox{Qatar-1b},
with uncertainties in the range of 24-70 seconds.
The linear ephemeris for both systems was improved. We couldn't determine any significant signal from O-C times of \mbox{TrES-3b};
however nearly clear periodicity of \mbox{\astrobj{Qatar-1b}} TTV was obtained at a 6.5\% significance level compared with
the FAP value of 28\% for the strongest peak given by \citet{Maciejewski2015AAP.577.109}.
\citet{vonEssen2013AAP.555.92} reported TTV periods of \mbox{$187 \pm 17$} and \mbox{$386 \pm 54$} days
with a fitted amplitude of \mbox{$A_{ttv} = 0.00052 \pm 0.00020$} for first peak.
These values are propotional with our TTV period of \mbox{$83.75 \pm 0.48$} days.
We estimated a TTV amplitude of \mbox{$A_{ttv} = 0.00044 \pm 0.00011$}.
This results give us support to result in that there is a perturbing object in the system.
After sinusoidal fit to O-C, we improved $\chi^{2}_{red}$ from 4.5 to 3.9.
The reason that we still have a high $\chi^{2}_{red}$ value, is that the data is spreading out a period of 8 years
with \mbox{$\sigma_{O-C} = 69.1$ seconds}. We had lowest individual points error of \mbox{$\sigma_{O} = 16.4$ seconds}
in O-C. At these conditions, to check the existence of new planets in the system,
more precise transit times spread over time are needed.
As given in the first paragraph of Section 5, time accuracy should be in the order of several tens of
seconds at least and data should spread over time in order to discover additional planets in star-planet systems
\citep[e.g.][]{Lithwick2012APJ.761.122}.

\begin{figure}
	\centering
	\includegraphics[width=11cm]{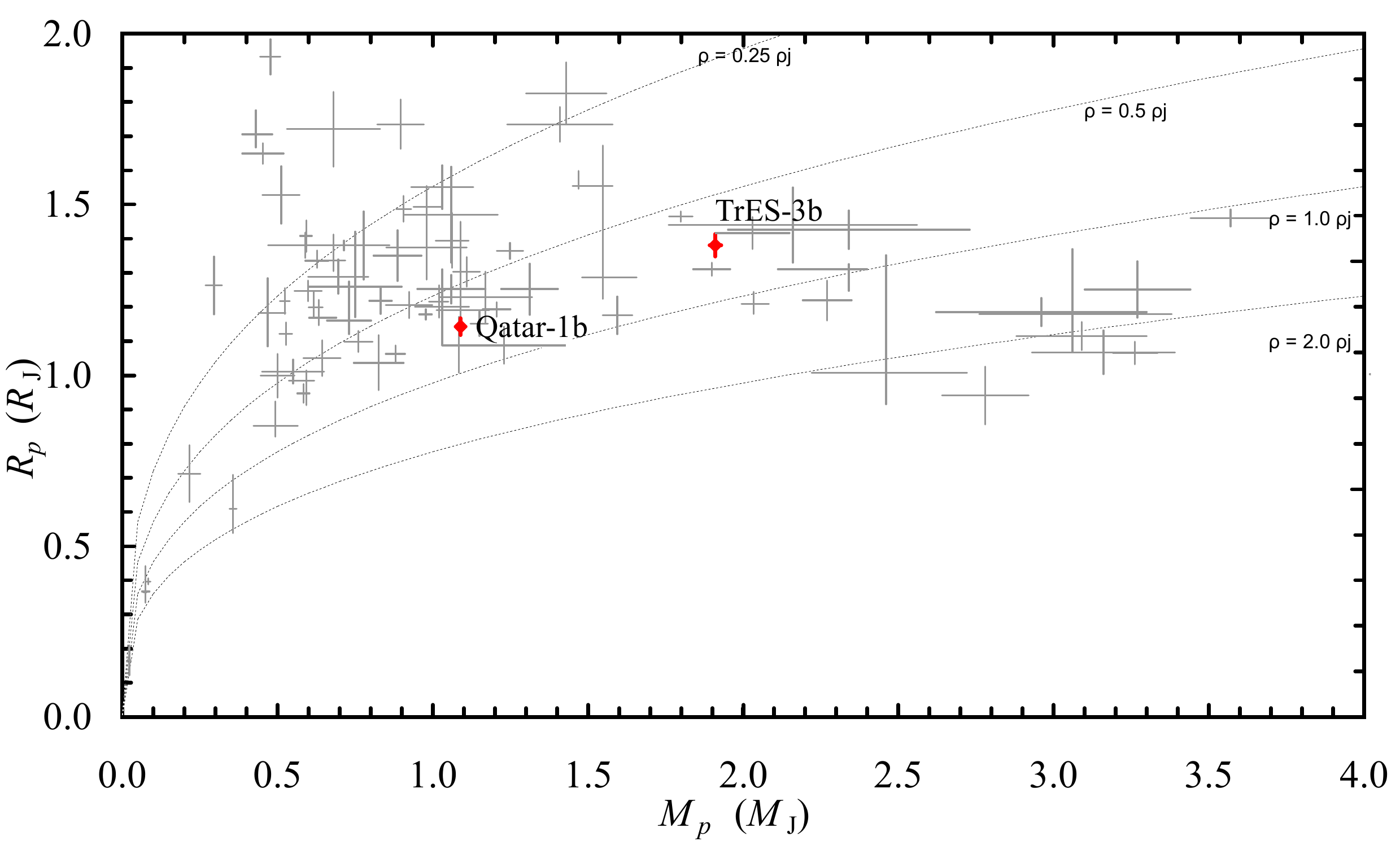}
    	\caption{Mass-radius relation of planets. Data apart from measurements of \mbox{TrES-3b} and \mbox{Qatar-1b} (red dots with errors) in this study were obtained from TEP catalogue.
	Dashed curves, top to bottom, show density, \mbox{$\rho = 0.25, 0.5, 1.0 $ and $ 2.0~\rho_{J}$}}
    \label{fig10}
\end{figure}

We collected data from the TEPCat catalogue for exoplanets having orbital radii smaller than $a = 0.024$ AU to
compare planets in general with our targets. As seen in Fig.~\ref{fig9}, WASP-46b has a similar orbit size and temperature
to that of \mbox{TrES-3b}, while there are only two known exoplanets, Qatar-2b and OGLE-TR-113b, with similar temperatures,
which are closer than \mbox{Qatar-1b} to their host stars. From the mass-radius relation of the planets presented in
Fig.~\ref{fig10}, one of our massive planets, \mbox{TrES-3b}, remains near the $\rho = 2.0 \rho_{J}$ density curve,
and appears rather exceptional in that respect. \mbox{Qatar-1b}, on the other hand, is more similar to Jupiter.

The stellar metallicities, \mbox{[Fe/H]\textsubscript{TrES-3} = $-0.2 <$ [Fe/H]\textsubscript{\mbox{Qatar-1}} = 0.2}
have raised some questions about the structure of the corresponding planets and their heating mechanisms
in the literature \citep{Torres2008APJ.677.1324, Alsubai2011MNRAS.417.709}.
The host star of \mbox{TrES-3b} may have magnetic activity, as mentioned by \citet{Christiansen2011APJ.726.94}.
While there is no signal of this in our data, it might be investigated again appending much more precise observations
extending further in time.

\section*{Acknowledgements}
This study was supported by T\"{U}B\.{I}TAK (Scientific and Technological Research Council of Turkey) under Grant No. 113F353.
We thank \c{C}anakkale Onsekiz Mart University Astrophysics Research Centre and Ulup{\i}nar Observatory and
\.{I}stanbul University Observatory Research and Application Center for their support and allowing use of T122 and T60
which were supported partly by National Planning Agency (DPT) of Turkey (project DPT-2007K120660 carried out \c{C}anakkale Onsekiz Mart University) and
the Scientific Research Projects Coordination Unit of \.{I}stanbul University (project no. 3685).
We thank to T\"{U}B\.{I}TAK for a partial support in using T100 telescope with project numbers 13CT100-523 and 13CT100-537.
This paper is part of the Ph.D. thesis of \c{C}.P\"{u}sk\"{u}ll\"{u}. 
This research made use of VIZIER and SIMBAD databases at CDS, Strasbourg, France.
We also thank TRESCA data providers and provider's observatories. In this respect, 
we thank Marc Bretton from Baronnies Provencales Observatory, 
Chen Cao from Weihai Observatory,  	
Christophe Gillier \& Romain Montaigut from Club Astronomie Lyon Ampere, 
Erika Pakstiene \& Rimvydas Janulis from VU Institute of Theoretical Physics and Astronomy, 
Paul Benni from Acton Sky Portal Observatory, 
Ken Hose from Quarryview Observatory, 
Juanjo Gonzalez from Cielo Profundo Observatory, 
Petri Kehusmaa \& Caisey Harlingten from Searchlight Observatory Network, 
Oleg Mazurenko from Trottier Observatory, 
Gajdos Stefan \& Ivana Jaksova from Modra Observatory 
and Marc Deldem from Les Barres Observatory.

\bibliographystyle{elsarticle-harv}
\bibliography{manuscript_references}

\begin{thebibliography}{36}
\expandafter\ifx\csname natexlab\endcsname\relax\def\natexlab#1{#1}\fi
\expandafter\ifx\csname url\endcsname\relax
  \def\url#1{\texttt{#1}}\fi
\expandafter\ifx\csname urlprefix\endcsname\relax\def\urlprefix{URL }\fi

\bibitem[{{Alsubai} et~al.(2011){Alsubai}, {Parley}, {Bramich}, {West},
  {Sorensen}, {Collier Cameron}, {Latham}, {Horne}, {Anderson}, {Bakos},
  {Brown}, {Buchhave}, {Esquerdo}, {Everett}, {F{\.z}r{\'e}sz}, {Hartman},
  {Hellier}, {Miller}, {Pollacco}, {Quinn}, {Smith}, {Stefanik}, and
  {Szentgyorgyi}}]{Alsubai2011MNRAS.417.709}
{Alsubai}, K.~A., {Parley}, N.~R., {Bramich}, D.~M., {West}, R.~G., {Sorensen},
  P.~M., {Collier Cameron}, A., {Latham}, D.~W., {Horne}, K., {Anderson},
  D.~R., {Bakos}, G.~{\'A}., {Brown}, D.~J.~A., {Buchhave}, L.~A., {Esquerdo},
  G.~A., {Everett}, M.~E., {F{\.z}r{\'e}sz}, G., {Hartman}, J.~D., {Hellier},
  C., {Miller}, G.~M., {Pollacco}, D., {Quinn}, S.~N., {Smith}, J.~C.,
  {Stefanik}, R.~P., {Szentgyorgyi}, A., Oct. 2011. {Q}atar-1b: a hot {J}upiter
  orbiting a metal-rich {K} dwarf star. MNRAS 417, 709--716.

\bibitem[{{Budding} and {Najim}(1980)}]{Budding1980APSS.72.369}
{Budding}, E., {Najim}, N.~N., Oct. 1980. {T}he system {VV} {ORI} and the
  consistency of photometric analysis of eclipsing binary light curves. Ap\&SS
  72, 369--396.

\bibitem[{{Budding} and {Zeilik}(1987)}]{Budding1987APJ.319.827}
{Budding}, E., {Zeilik}, M., Aug. 1987. {A}n analysis of the light curves of
  short-period {RS} {C}anum {V}enaticorum stars - {S}tarspots and fundamental
  properties. ApJ 319, 827--835.

\bibitem[{{Christiansen} et~al.(2011){Christiansen}, {Ballard}, {Charbonneau},
  {Deming}, {Holman}, {Madhusudhan}, {Seager}, {Wellnitz}, {Barry},
  {Livengood}, {Hewagama}, {Hampton}, {Lisse}, and
  {A'Hearn}}]{Christiansen2011APJ.726.94}
{Christiansen}, J.~L., {Ballard}, S., {Charbonneau}, D., {Deming}, D.,
  {Holman}, M.~J., {Madhusudhan}, N., {Seager}, S., {Wellnitz}, D.~D., {Barry},
  R.~K., {Livengood}, T.~A., {Hewagama}, T., {Hampton}, D.~L., {Lisse}, C.~M.,
  {A'Hearn}, M.~F., Jan. 2011. {S}ystem {P}arameters, {T}ransit {T}imes, and
  {S}econdary {E}clipse {C}onstraints of the {E}xoplanet {S}ystems {HAT}-{P}-4,
  {T}r{ES}-2, {T}r{ES}-3, and {WASP}-3 from the {NASA} {EPOXI} {M}ission of
  {O}pportunity. ApJ 726, 94.

\bibitem[{{Ciceri} et~al.(2015){Ciceri}, {Mancini}, {Southworth}, {Bruni},
  {Nikolov}, {D'Ago}, {Schr{\"o}der}, {Bozza}, {Tregloan-Reed}, and
  {Henning}}]{Ciceri2015AAP.577.54}
{Ciceri}, S., {Mancini}, L., {Southworth}, J., {Bruni}, I., {Nikolov}, N.,
  {D'Ago}, G., {Schr{\"o}der}, T., {Bozza}, V., {Tregloan-Reed}, J., {Henning},
  T., May 2015. {P}hysical properties of the {HAT}-{P}-23 and {WASP}-48
  planetary systems from multi-colour photometry. A\&A 577, A54.

\bibitem[{{Claret} and {Bloemen}(2011)}]{Claret2011AAP.529.75}
{Claret}, A., {Bloemen}, S., May 2011. {G}ravity and limb-darkening
  coefficients for the {K}epler, {C}o{R}o{T}, {S}pitzer, uvby, {UBVRIJHK}, and
  {S}loan photometric systems. A\&A 529, A75.

\bibitem[{{Collins} et~al.(2015){Collins}, {Kielkopf}, and
  {Stassun}}]{Collins2015ARXIV}
{Collins}, K.~A., {Kielkopf}, J.~F., {Stassun}, K.~G., Dec. 2015. {T}ransit
  {T}iming {V}ariation {M}easurements of {WASP}-12b and {Q}atar-1b: {N}o
  {E}vidence for {A}dditional {P}lanets. arXiv.

\bibitem[{{Col{\'o}n} et~al.(2010){Col{\'o}n}, {Ford}, {Lee}, {Mahadevan}, and
  {Blake}}]{Colon2010MNRAS.408.1494}
{Col{\'o}n}, K.~D., {Ford}, E.~B., {Lee}, B., {Mahadevan}, S., {Blake}, C.~H.,
  Nov. 2010. {C}haracterizing transiting extrasolar planets with narrow-band
  photometry and {GTC}/{OSIRIS}. MNRAS 408, 1494--1501.

\bibitem[{{Covino} et~al.(2013){Covino}, {Esposito}, {Barbieri}, {Mancini},
  {Nascimbeni}, {Claudi}, {Desidera}, {Gratton}, {Lanza}, {Sozzetti}, {Biazzo},
  {Affer}, {Gandolfi}, {Munari}, {Pagano}, {Bonomo}, {Collier Cameron},
  {H{\'e}brard}, {Maggio}, {Messina}, {Micela}, {Molinari}, {Pepe}, {Piotto},
  {Ribas}, {Santos}, {Southworth}, {Shkolnik}, {Triaud}, {Bedin}, {Benatti},
  {Boccato}, {Bonavita}, {Borsa}, {Borsato}, {Brown}, {Carolo}, {Ciceri},
  {Cosentino}, {Damasso}, {Faedi}, {Mart{\'{\i}}nez Fiorenzano}, {Latham},
  {Lovis}, {Mordasini}, {Nikolov}, {Poretti}, {Rainer}, {Rebolo L{\'o}pez},
  {Scandariato}, {Silvotti}, {Smareglia}, {Alcal{\'a}}, {Cunial}, {Di
  Fabrizio}, {Di Mauro}, {Giacobbe}, {Granata}, {Harutyunyan}, {Knapic},
  {Lattanzi}, {Leto}, {Lodato}, {Malavolta}, {Marzari}, {Molinaro},
  {Nardiello}, {Pedani}, {Prisinzano}, and {Turrini}}]{Covino2013AAP.554.28}
{Covino}, E., {Esposito}, M., {Barbieri}, M., {Mancini}, L., {Nascimbeni}, V.,
  {Claudi}, R., {Desidera}, S., {Gratton}, R., {Lanza}, A.~F., {Sozzetti}, A.,
  {Biazzo}, K., {Affer}, L., {Gandolfi}, D., {Munari}, U., {Pagano}, I.,
  {Bonomo}, A.~S., {Collier Cameron}, A., {H{\'e}brard}, G., {Maggio}, A.,
  {Messina}, S., {Micela}, G., {Molinari}, E., {Pepe}, F., {Piotto}, G.,
  {Ribas}, I., {Santos}, N.~C., {Southworth}, J., {Shkolnik}, E., {Triaud},
  A.~H.~M.~J., {Bedin}, L., {Benatti}, S., {Boccato}, C., {Bonavita}, M.,
  {Borsa}, F., {Borsato}, L., {Brown}, D., {Carolo}, E., {Ciceri}, S.,
  {Cosentino}, R., {Damasso}, M., {Faedi}, F., {Mart{\'{\i}}nez Fiorenzano},
  A.~F., {Latham}, D.~W., {Lovis}, C., {Mordasini}, C., {Nikolov}, N.,
  {Poretti}, E., {Rainer}, M., {Rebolo L{\'o}pez}, R., {Scandariato}, G.,
  {Silvotti}, R., {Smareglia}, R., {Alcal{\'a}}, J.~M., {Cunial}, A., {Di
  Fabrizio}, L., {Di Mauro}, M.~P., {Giacobbe}, P., {Granata}, V.,
  {Harutyunyan}, A., {Knapic}, C., {Lattanzi}, M., {Leto}, G., {Lodato}, G.,
  {Malavolta}, L., {Marzari}, F., {Molinaro}, M., {Nardiello}, D., {Pedani},
  M., {Prisinzano}, L., {Turrini}, D., Jun. 2013. {T}he {GAPS} programme with
  {HARPS}-{N} at {TNG}. {I}. {O}bservations of the {R}ossiter-{M}c{L}aughlin
  effect and characterisation of the transiting system {Q}atar-1. A\&A 554,
  A28.

\bibitem[{{Eastman} et~al.(2010){Eastman}, {Siverd}, and
  {Gaudi}}]{Eastman2010PASP.122.935}
{Eastman}, J., {Siverd}, R., {Gaudi}, B.~S., Aug. 2010. {A}chieving {B}etter
  {T}han 1 {M}inute {A}ccuracy in the {H}eliocentric and {B}arycentric {J}ulian
  {D}ates. PASP 122, 935--946.

\bibitem[{{Fressin} et~al.(2010){Fressin}, {Knutson}, {Charbonneau},
  {O'Donovan}, {Burrows}, {Deming}, {Mandushev}, and
  {Spiegel}}]{Fressin2010APJ.711.374}
{Fressin}, F., {Knutson}, H.~A., {Charbonneau}, D., {O'Donovan}, F.~T.,
  {Burrows}, A., {Deming}, D., {Mandushev}, G., {Spiegel}, D., Mar. 2010. {T}he
  {B}roadband {I}nfrared {E}mission {S}pectrum of the {E}xoplanet {T}r{ES}-3.
  ApJ 711, 374--379.

\bibitem[{{Gibson} et~al.(2009){Gibson}, {Pollacco}, {Simpson}, {Barros},
  {Joshi}, {Todd}, {Keenan}, {Skillen}, {Benn}, {Christian}, {Hrudkov{\'a}},
  and {Steele}}]{Gibson2009APJ.700.1078}
{Gibson}, N.~P., {Pollacco}, D., {Simpson}, E.~K., {Barros}, S., {Joshi},
  Y.~C., {Todd}, I., {Keenan}, F.~P., {Skillen}, I., {Benn}, C., {Christian},
  D., {Hrudkov{\'a}}, M., {Steele}, I.~A., Aug. 2009. {A} {T}ransit {T}iming
  {A}nalysis of {N}ine {R}ise {L}ight {C}urves of the {E}xoplanet {S}ystem
  {T}r{ES}-3. ApJ 700, 1078--1085.

\bibitem[{{Hansen} and {Barman}(2007)}]{Hansen2007APJ.671.861}
{Hansen}, B.~M.~S., {Barman}, T., Dec. 2007. {T}wo {C}lasses of {H}ot
  {J}upiters. ApJ 671, 861--871.

\bibitem[{{Horne} and {Baliunas}(1986)}]{Horne1986APJ.302.757}
{Horne}, J.~H., {Baliunas}, S.~L., Mar. 1986. {A} prescription for period
  analysis of unevenly sampled time series. ApJ 302, 757--763.

\bibitem[{{Kopal}(1959)}]{Kopal1959}
{Kopal}, Z., 1959. {C}lose binary systems. Chapman \& Hall, London.

\bibitem[{{Kundurthy} et~al.(2013){Kundurthy}, {Becker}, {Agol}, {Barnes}, and
  {Williams}}]{Kundurthy2013APJ.764.8}
{Kundurthy}, P., {Becker}, A.~C., {Agol}, E., {Barnes}, R., {Williams}, B.,
  Feb. 2013. {APOSTLE}: 11 {T}ransit {O}bservations of {T}r{ES}-3b. ApJ 764, 8.

\bibitem[{{Lee} et~al.(2011){Lee}, {Youn}, {Kim}, {Lee}, and
  {Koo}}]{Lee2011PASJ.63.301}
{Lee}, J.~W., {Youn}, J.-H., {Kim}, S.-L., {Lee}, C.-U., {Koo}, J.-R., Feb.
  2011. {P}hysical {P}roperties of the {T}ransiting {P}lanetary {S}ystem
  {T}r{ES}-3. PASJ 63, 301.

\bibitem[{Lithwick et~al.(2012)Lithwick, Xie, and Wu}]{Lithwick2012APJ.761.122}
Lithwick, Y., Xie, J., Wu, Y., Dec. 2012. {Extracting Planet Mass and
  Eccentricity from TTV Data}. ApJ 761, 122.

\bibitem[{{Lomb}(1976)}]{Lomb1976APSS.39.447}
{Lomb}, N.~R., Feb. 1976. {L}east-squares frequency analysis of unequally
  spaced data. Ap\&SS 39, 447--462.

\bibitem[{{Maciejewski} et~al.(2015){Maciejewski}, {Fern{\'a}ndez}, {Aceituno},
  {Ohlert}, {Puchalski}, {Dimitrov}, {Seeliger}, {Kitze}, {Raetz}, {Errmann},
  {Gilbert}, {Pannicke}, {Schmidt}, and
  {Neuh{\"a}user}}]{Maciejewski2015AAP.577.109}
{Maciejewski}, G., {Fern{\'a}ndez}, M., {Aceituno}, F.~J., {Ohlert}, J.,
  {Puchalski}, D., {Dimitrov}, D., {Seeliger}, M., {Kitze}, M., {Raetz}, S.,
  {Errmann}, R., {Gilbert}, H., {Pannicke}, A., {Schmidt}, J.-G.,
  {Neuh{\"a}user}, R., May 2015. {N}o variations in transit times for {Q}atar-1
  b. A\&A 577, A109.

\bibitem[{{Mislis} et~al.(2015){Mislis}, {Mancini}, {Tregloan-Reed}, {Ciceri},
  {Southworth}, {D'Ago}, {Bruni}, {Ba{\c s}t{\"u}rk}, {Alsubai}, {Bachelet},
  {Bramich}, {Henning}, {Hinse}, {Iannella}, {Parley}, and
  {Schroeder}}]{Mislis2015MNRAS.448.2617}
{Mislis}, D., {Mancini}, L., {Tregloan-Reed}, J., {Ciceri}, S., {Southworth},
  J., {D'Ago}, G., {Bruni}, I., {Ba{\c s}t{\"u}rk}, {\"O}., {Alsubai}, K.~A.,
  {Bachelet}, E., {Bramich}, D.~M., {Henning}, T., {Hinse}, T.~C., {Iannella},
  A.~L., {Parley}, N., {Schroeder}, T., Apr. 2015. {H}igh-precision multiband
  time series photometry of exoplanets {Q}atar-1b and {T}r{ES}-5b. MNRAS 448,
  2617--2623.

\bibitem[{{O'Donovan} et~al.(2007){O'Donovan}, {Charbonneau}, {Bakos},
  {Mandushev}, {Dunham}, {Brown}, {Latham}, {Torres}, {Sozzetti}, {Kov{\'a}cs},
  {Everett}, {Baliber}, {Hidas}, {Esquerdo}, {Rabus}, {Deeg}, {Belmonte},
  {Hillenbrand}, and {Stefanik}}]{O'Donovan2007APJ.663.37}
{O'Donovan}, F.~T., {Charbonneau}, D., {Bakos}, G.~{\'A}., {Mandushev}, G.,
  {Dunham}, E.~W., {Brown}, T.~M., {Latham}, D.~W., {Torres}, G., {Sozzetti},
  A., {Kov{\'a}cs}, G., {Everett}, M.~E., {Baliber}, N., {Hidas}, M.~G.,
  {Esquerdo}, G.~A., {Rabus}, M., {Deeg}, H.~J., {Belmonte}, J.~A.,
  {Hillenbrand}, L.~A., {Stefanik}, R.~P., Jul. 2007. {T}r{ES}-3: {A} {N}earby,
  {M}assive, {T}ransiting {H}ot {J}upiter in a 31 {H}our {O}rbit. ApJ 663,
  L37--L40.

\bibitem[{{Petrucci} et~al.(2015){Petrucci}, {Jofr{\'e}}, {Melita},
  {G{\'o}mez}, and {Mauas}}]{Petrucci2015MNRAS.446.1389}
{Petrucci}, R., {Jofr{\'e}}, E., {Melita}, M., {G{\'o}mez}, M., {Mauas}, P.,
  Jan. 2015. {T}ransit timing variation analysis in southern stars: the case of
  {WASP}-28. MNRAS 446, 1389--1398.

\bibitem[{{Poddan{\'y}} et~al.(2010){Poddan{\'y}}, {Br{\'a}t}, and
  {Pejcha}}]{Poddany2010NA.15.297}
{Poddan{\'y}}, S., {Br{\'a}t}, L., {Pejcha}, O., Mar. 2010. {Exoplanet Transit
  Database. Reduction and processing of the photometric data of exoplanet
  transits}. NewA 15, 297--301.

\bibitem[{{Rhodes} and {Budding}(2014)}]{Rhodes2014APSS.351.451}
{Rhodes}, M.~D., {Budding}, E., Jun. 2014. {A}nalysis of selected {K}epler
  {M}ission planetary light curves. Ap\&SS 351, 451--471.

\bibitem[{{Scargle}(1982)}]{Scargle1982APJ.263.835}
{Scargle}, J.~D., Dec. 1982. {S}tudies in astronomical time series analysis.
  {II} - {S}tatistical aspects of spectral analysis of unevenly spaced data.
  ApJ 263, 835--853.

\bibitem[{{Southworth}(2008)}]{Southworth2008MNRAS.386.1644}
{Southworth}, J., May 2008. {H}omogeneous studies of transiting extrasolar
  planets - {I}. {L}ight-curve analyses. MNRAS 386, 1644--1666.

\bibitem[{{Southworth}(2010)}]{Southworth2010MNRAS.408.1689}
{Southworth}, J., Nov. 2010. {H}omogeneous studies of transiting extrasolar
  planets - {III}. {A}dditional planets and stellar models. MNRAS 408,
  1689--1713.

\bibitem[{{Southworth}(2011)}]{Southworth2011MNRAS.417.2166}
{Southworth}, J., Nov. 2011. {H}omogeneous studies of transiting extrasolar
  planets - {IV}. {T}hirty systems with space-based light curves. MNRAS 417,
  2166--2196.

\bibitem[{{Southworth}(2012)}]{Southworth2012.282.131}
{Southworth}, J., Apr. 2012. {H}omogeneous {S}tudies of {T}ransiting {P}lanets.
  In: {Richards}, M.~T., {Hubeny}, I. (Eds.), From Interacting Binaries to
  Exoplanets: Essential Modeling Tools. Vol. 282 of IAU Symposium. pp.
  131--132.

\bibitem[{{Sozzetti} et~al.(2009){Sozzetti}, {Torres}, {Charbonneau}, {Winn},
  {Korzennik}, {Holman}, {Latham}, {Laird}, {Fernandez}, {O'Donovan},
  {Mandushev}, {Dunham}, {Everett}, {Esquerdo}, {Rabus}, {Belmonte}, {Deeg},
  {Brown}, {Hidas}, and {Baliber}}]{Sozzetti2009APJ.691.1145}
{Sozzetti}, A., {Torres}, G., {Charbonneau}, D., {Winn}, J.~N., {Korzennik},
  S.~G., {Holman}, M.~J., {Latham}, D.~W., {Laird}, J.~B., {Fernandez}, J.,
  {O'Donovan}, F.~T., {Mandushev}, G., {Dunham}, E., {Everett}, M.~E.,
  {Esquerdo}, G.~A., {Rabus}, M., {Belmonte}, J.~A., {Deeg}, H.~J., {Brown},
  T.~N., {Hidas}, M.~G., {Baliber}, N., Feb. 2009. {A} {N}ew {S}pectroscopic
  and {P}hotometric {A}nalysis of the {T}ransiting {P}lanet {S}ystems
  {T}r{ES}-3 and {T}r{ES}-4. ApJ 691, 1145--1158.

\bibitem[{{Torres} et~al.(2008){Torres}, {Winn}, and
  {Holman}}]{Torres2008APJ.677.1324}
{Torres}, G., {Winn}, J.~N., {Holman}, M.~J., Apr. 2008. {I}mproved
  {P}arameters for {E}xtrasolar {T}ransiting {P}lanets. ApJ 677, 1324--1342.

\bibitem[{{Turner} et~al.(2013){Turner}, {Smart}, {Hardegree-Ullman},
  {Carleton}, {Walker-LaFollette}, {Crawford}, {Smith}, {McGraw}, {Small},
  {Rocchetto}, {Cunningham}, {Towner}, {Zellem}, {Robertson}, {Guvenen},
  {Schwarz}, {Hardegree-Ullman}, {Collura}, {Henz}, {Lejoly}, {Richardson},
  {Weinand}, {Taylor}, {Daugherty}, {Wilson}, and
  {Austin}}]{Turner2013MNRAS.428.678}
{Turner}, J.~D., {Smart}, B.~M., {Hardegree-Ullman}, K.~K., {Carleton}, T.~M.,
  {Walker-LaFollette}, A.~M., {Crawford}, B.~E., {Smith}, C.-T.~W., {McGraw},
  A.~M., {Small}, L.~C., {Rocchetto}, M., {Cunningham}, K.~I., {Towner},
  A.~P.~M., {Zellem}, R., {Robertson}, A.~N., {Guvenen}, B.~C., {Schwarz},
  K.~R., {Hardegree-Ullman}, E.~E., {Collura}, D., {Henz}, T.~N., {Lejoly}, C.,
  {Richardson}, L.~L., {Weinand}, M.~A., {Taylor}, J.~M., {Daugherty}, M.~J.,
  {Wilson}, A.~A., {Austin}, C.~L., Jan. 2013. {N}ear-{UV} and optical
  observations of the transiting exoplanet {T}r{ES}-3b. MNRAS 428, 678--690.

\bibitem[{{Va{\v n}ko} et~al.(2013){Va{\v n}ko}, {Maciejewski},
  {Jakub{\'{\i}}k}, {Krej{\v c}ov{\'a}}, {Budaj}, {Pribulla}, {Ohlert},
  {Raetz}, {Parimucha}, and {Bukowiecki}}]{Vavnko2013MNRAS.432.944}
{Va{\v n}ko}, M., {Maciejewski}, G., {Jakub{\'{\i}}k}, M., {Krej{\v c}ov{\'a}},
  T., {Budaj}, J., {Pribulla}, T., {Ohlert}, J., {Raetz}, S., {Parimucha}, {\v
  S}., {Bukowiecki}, L., Jun. 2013. {P}hotometric follow-up of the transiting
  planetary system {T}r{ES}-3: transit timing variation and long-term stability
  of the system. MNRAS 432, 944--953.

\bibitem[{{von Essen} et~al.(2013){von Essen}, {Schr{\"o}ter}, {Agol}, and
  {Schmitt}}]{vonEssen2013AAP.555.92}
{von Essen}, C., {Schr{\"o}ter}, S., {Agol}, E., {Schmitt}, J.~H.~M.~M., Jul.
  2013. {Q}atar-1: indications for possible transit timing variations. A\&A
  555, A92.

\bibitem[{{Winn} et~al.(2008){Winn}, {Holman}, {Torres}, {McCullough},
  {Johns-Krull}, {Latham}, {Shporer}, {Mazeh}, {Garcia-Melendo}, {Foote},
  {Esquerdo}, and {Everett}}]{Winn2008APJ.683.1076}
{Winn}, J.~N., {Holman}, M.~J., {Torres}, G., {McCullough}, P., {Johns-Krull},
  C., {Latham}, D.~W., {Shporer}, A., {Mazeh}, T., {Garcia-Melendo}, E.,
  {Foote}, C., {Esquerdo}, G., {Everett}, M., Aug. 2008. {T}he {T}ransit
  {L}ight {C}urve {P}roject. {IX}. {E}vidence for a {S}maller {R}adius of the
  {E}xoplanet {XO}-3b. ApJ 683, 1076--1084.

\end{thebibliography}



\label{lastpage}
\end{document}